\documentclass[traditabstract]{aa} 

\usepackage{graphicx}
\usepackage{txfonts}
\usepackage{longtable}
\begin{document}
   \title{Spectacular tails of ionised gas in the Virgo cluster galaxy NGC 4569\thanks{Based on observations obtained with
   MegaPrime/MegaCam, a joint project of CFHT and CEA/DAPNIA, at the Canadian-French-Hawaii Telescope
   (CFHT) which is operated by the National Research Council (NRC) of Canada, the Institut National
   des Sciences de l'Univers of the Centre National de la Recherche Scientifique (CNRS) of France and
   the University of Hawaii.}}
   \subtitle{}
  \author{A. Boselli\inst{1},         
	  J.C. Cuillandre\inst{2},
	  M. Fossati\inst{3,4},	  
	  S. Boissier\inst{1},
	  D. Bomans\inst{5},
	  G. Consolandi\inst{6},	  
	  G. Anselmi\inst{7},	  
	  L. Cortese\inst{8},	
	  P. C{\^o}t{\'e}\inst{9},
	  P. Durrell\inst{10},
	  L. Ferrarese\inst{9},
	  M. Fumagalli\inst{11},  
	  G. Gavazzi\inst{6},
	  S. Gwyn\inst{9},	
	  G. Hensler\inst{12,13},
	  M. Sun\inst{14},
	  E. Toloba\inst{15,16}
       }

\institute{	
		Aix Marseille Universit\'e, CNRS, LAM (Laboratoire d'Astrophysique de Marseille), UMR 7326, F-13388, Marseille, France
             \email{alessandro.boselli@lam.fr, samuel.boissier@lam.fr}
        \and  
		CEA/IRFU/SAP, Laboratoire AIM Paris-Saclay, CNRS/INSU, Université Paris Diderot, Observatoire de Paris, PSL Research University, F-91191 Gif-sur-Yvette Cedex, France
		\email{jc.cuillandre@cea.fr}
	\and
	        Universit{\"a}ts-Sternwarte M{\"u}nchen, Scheinerstrasse 1, D-81679 M{\"u}nchen, Germany
        \and
                Max-Planck-Institut f\"{u}r Extraterrestrische Physik, Giessenbachstrasse, 85748, Garching, Germany 
                \email{mfossati@mpe.mpg.de}
	\and
		Astronomical Institute of the Ruhr-Universit\"at Bochum, Universit\"atsstr. 150, 44801 Bochum, Germany
		\email{bomans@astro.rub.de}
	\and
	 	Universit\'a di Milano-Bicocca, piazza della scienza 3, 20100, Milano, Italy
		\email{guido.consolandi@mib.infn.it, giuseppe.gavazzi@mib.infn.it}
	\and	
		Coelum Astronomia, via Appia 20, 30173 Venezia, Italy
		\email{giovanni.anselimi@coelum.com}
	\and	
		International Centre for Radio Astronomy Research, The University of Western Australia, 35 Stirling Highway, Crawley WA 6009, Australia
		\email{luca.cortese@uwa.edu.au}
	\and
		NRC Herzberg Astronomy and Astrophysics, 5071 West Saanich Road, Victoria, BC, V9E 2E7, Canada
		\email{laura.ferrarese@nrc-cnrc.gr.ca, patrick.cote@nrc-cnrc.gr.ca, stephen.gwyn@nrc-cnrc.gr.ca}
	\and
		Department of Physiscs and Astronomy, Youngstown State University, Youngstown, OH, USA
		\email{prdurrel@ysu.edu}
	\and
		Institute for Computational Cosmology and Centre for Extragalactic Astronomy, Department of Physics, Durham University, South Road, Durham DH1 3LE, UK
 		\email{michele.fumagalli@durham.ac.uk}
	\and
		Department of Astrophysics, University of Vienna, Türkenschanzstrasse 17, 1180, Vienna, Austria
		\email{gerhard.hensler@univie.ac.at}
	\and
		National Astronomy Observatory of Japan, 2-21-1 Osawa, Mitaka-shi, Tokyo 181-8588, Japan
	\and
		Physics Department, University of Alabama in Huntsville, Huntsville, AL 35899, USA
		\email{ms0071@uah.edu}
	\and
		UCO/Lick Observatory, University of California, Santa Cruz, 1156 High Street, Santa Cruz, CA 95064, USA
		\email{toloba@ucolik.org}
	\and
		Texas Tech University, Physics Department, Box 41051, Lubbock, TX 79409-1051, USA
 		}
               
\authorrunning{Boselli et al.}
\titlerunning{Spectacular tails of ionised gas in the Virgo cluster galaxy NGC 4569}

   \date{}

 
  \abstract  
{We obtained using MegaCam at the CFHT a deep narrow band H$\alpha$+[NII] wide field image of NGC 4569 (M90), the brightest late-type galaxy in the Virgo cluster. The image
reveals the presence of long tails of diffuse ionised gas without any associated stellar component extending from the disc of the galaxy up to 
$\simeq$ 80 kpc (projected distance) with a typical surface brightness 
of a few 10$^{-18}$ erg s$^{-1}$ cm$^{-2}$ arcsec$^{-2}$. These features provide direct evidence that NGC 4569 is undergoing a ram presure stripping event.
The image also shows a prominent 8 kpc spur of ionised gas associated to the nucleus that spectroscopic data identify 
as an outflow. With some assumptions on the 3D distribution of the gas, we use the H$\alpha$ surface brightness of these extended low surface
brightness features to derive the density and the mass of the gas stripped during the interaction of the galaxy with the intracluster medium.
The comparison with ad-hoc chemo-spectrophotometric models of galaxy evolution indicates that the mass of the H$\alpha$ emitting gas in the tail is comparable to that 
of the cold phase stripped from the disc, suggesting that the gas is ionised within the tail during the stripping process. The lack of star forming regions
suggests that mechanisms other than photoionisation are responsible for the excitation of the gas (shocks, heat conduction, magneto hydrodynamic waves).
This analysis indicates that ram pressure stripping is efficient in massive ($M_{star}$ $\simeq$ 10$^{10.5}$ M$_{\odot}$) galaxies located in 
intermediate mass ($\simeq$ 10$^{14}$ M$_{\odot}$) clusters under formation. It also shows that the mass of gas expelled by the nuclear outflow is  
$\sim$ 1 \% than that removed during the ram pressure stripping event. All together these results indicate that ram pressure stripping, 
rather than starvation through nuclear feedback, can be the dominant mechanism responsible for the quenching of the star formation activity 
of galaxies in high density environments. 
 }
   {}
   {}
   {}
   {}
   {}


   \maketitle
%

\section{Introduction}

The environment plays a fundamental role in galaxy evolution. Since the early works of Hubble \& Humason (1931),
Abell (1965), and Oemler (1974) 
it became evident that galaxies in rich environments are systematically different from those located in the field. 
Quiescent objects (ellipticals and lenticulars) are dominating high-density regions such as clusters and compact groups, 
while late-type systems are mostly located in the field (Dressler 1980; Binggeli et al. 1988; 
Whitmore et al. 1993; Dressler et al. 1997). 
It also became clear that the physical properties of star forming systems 
inhabiting rich environments are systematically different from those of their isolated analogues, with a 
reduced atomic (e.g. Cayatte et al. 1990; Solanes et al. 2001; Vollmer et al. 2001; Gavazzi et al. 2006a) and molecular gas content (Fumagalli et al. 2009; 
Boselli et al. 2014c), dust content (Cortese et al. 2010; 2012a), and star formation 
(e.g. Kennicutt 1983; Gavazzi et al. 1998, 2006b, 2013; Lewis et al. 2002; 
Goto et al. 2003; Boselli et al. 2015). 
   \begin{figure*}
   \centering
   \includegraphics[width=18cm]{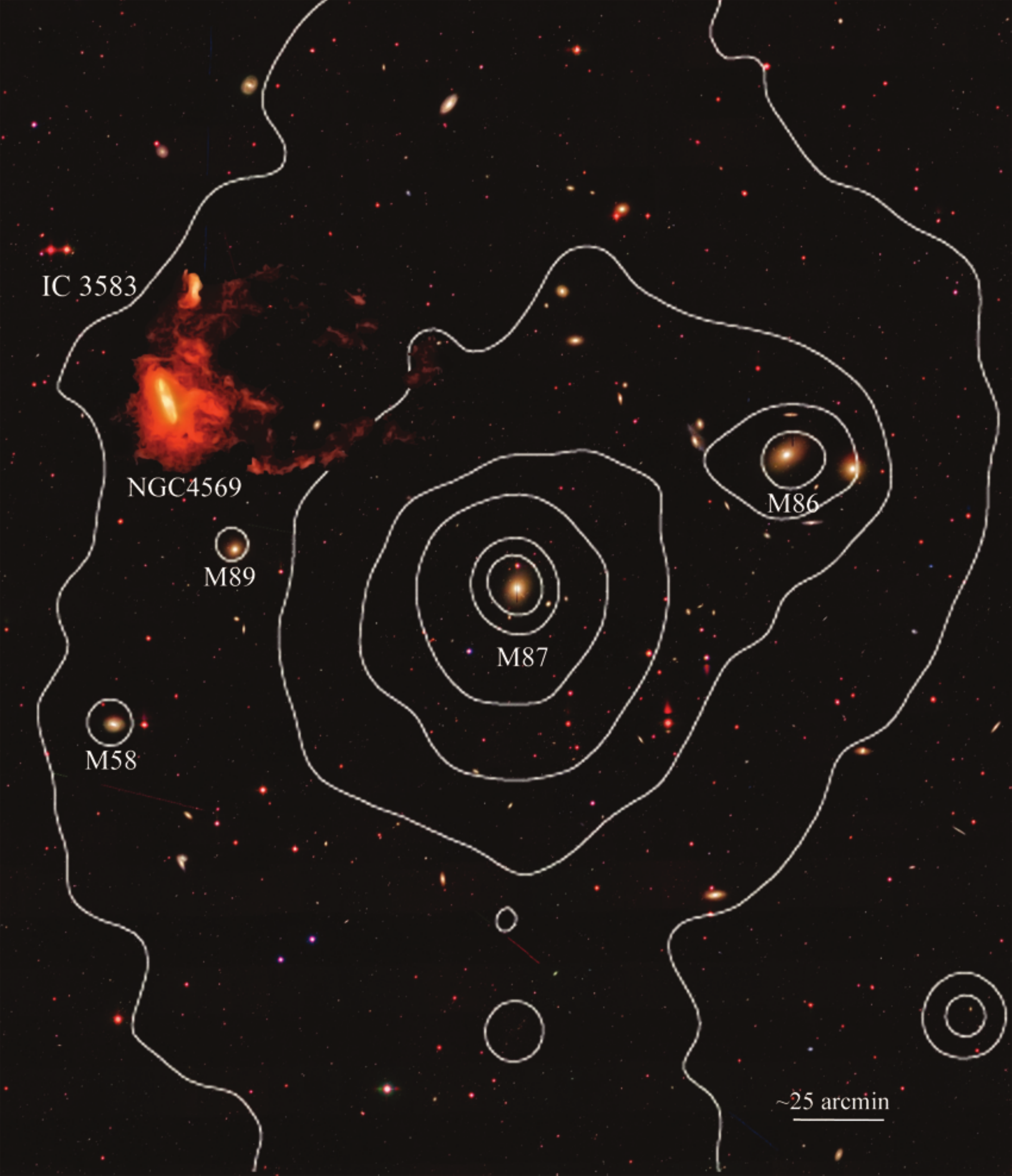}
   \caption{The galaxy NGC 4569 is located at 1.7 degrees (0.32 $R_{vir}$) north-east from the core of the Virgo cluster M87. Contours indicate
   the distribution of the X-ray gas as derived by ROSAT observations (B\"ohringer et al. 1994). The CFHT MegaCam continuum-subtracted 
   H$\alpha$+[NII] image of the galaxy smoothed with a 5x5 median filter
   and masked form the emission of foreground stars (see sect. 3) is magnified by a factor of 6 to underline the orientation of the tail of ionised gas with respect 
   to the position of the galaxy within the cluster.
 }
   \label{Virgo}%
   \end{figure*}

As reviewed in Boselli \& Gavazzi (2006), several physical mechanisms have been proposed to explain the origin of these differences. 
These processes belongs to two main families, those related to the gravitational interactions between galaxies or with the potential 
well of the overedense region (tidal interactions - Merritt 1983; Byrd \& Valtonen 1990, harassment - Moore et al. 1998), 
and those exerted by the hot and dense intracluster medium on galaxies moving at high velocity within the 
cluster (ram pressure stripping - Gunn \& Gott 1972, viscous stripping - Nulsen 1982, thermal evaporation 
- Cowie \& Songaila 1977, starvation - Larson et al. 1980). Since these large dynamically-bounded structures 
observed in the local universe form through the accretion of smaller groups of galaxies, 
these processes might start to act well before galaxies enter rich clusters  (pre-processing; Dressler 2004). 

The identification and the physical understanding of the dominant process affecting galaxies in rich 
environments at different epochs is fundamental for tuning cosmological models of galaxy evolution. 
At present, observations and simulations give discordant results whenever large statistical samples extracted from blind 
surveys are compared to targeted observations of nearby clusters and groups. 
Most hydrodynamic cosmological simulations suggest that the environmental quenching of the star formation 
is mainly regulated by starvation. Once in high-density regions, galaxies lose their hot gas halo during the dynamical interaction with the hostile
environment. AGN and supernovae feedback becomes particularly efficient in ejecting the gas 
out from the galactic disc, quenching on several Gyr the activity of star formation (McGee et al. 2009; Weinmann et al. 2010). 
This scenario is supported by the analysis of SDSS data, which suggest a quenching timescale of $\sim$ 5 Gyr for galaxies 
in dense environments (Wetzel et al. 2012, 2013). It is also supported by the observations of several clusters at 
intermediate redshift (e.g. Haines et al. 2013, 2015).
In its current form, however, this scenario over-predicts the fraction of red dwarf galaxies compared to what is observed 
in nearby clusters (Kang \& van den Bosch 2008; Font et al. 2008; Kimm et al. 2009; De Lucia 2011; Weinmann et al. 2011; Taranu et al. 2014). 
At the same time chemo-spectrophotometric multizone models of starvation fail to reproduce the observed radial profiles of the cold gas 
and of the young stars of star forming galaxies in nearby clusters (Boselli et al. 2006). 

Recent hydrodynamic simulations of individual galaxies indicate ram pressure as a compelling alternative process to explain the observed
peculiarities of cluster members (Roediger \& Bruggen 2007, 2008; Tonnesen \& Bryan 2009, 2010, 2012). These simulations show that, whenever the different gas 
phases are properly simulated at high resolution, ram pressure is the dominant mechanism responsible for the gas stripping and for the following quenching of
star formation up to $\sim$ 1 virial radius of the cluster. The recent observation of several 
late-type galaxies with long tails of gas without any associated stellar component at large
clustercentric distances seems to corroborate this scenario (e.g. Scott et al. 2012; Yagi et al. 2010; Fossati et al. 2012). However, it remains
unclear what is the contribution of the nuclear feedback to the stripping process, in particular in massive spiral galaxies 
where both the gravitational potential well and the nuclear activity are maximal. To date, the direct observations of the feedback process in clusters 
is still limited to a few central early-type galaxies with cooling flows (Fabian 2012). The detailed study of star forming systems is thus urgent
to understand the role of feedback in the environmental quenching of the star formation activity.

A full understanding of the gas stripping process in high density regions requires the
comparison of multifrequency observations covering the different phases of the interstellar medium (ISM; atomic and molecular, ionised, hot gas, dust)
and the different stellar populations with tuned chemo-spectrophotometric and hydrodynamic models of gas stripping.
This comparison has been done in the Virgo cluster (Boselli et al. 2014b), the closest concentration of galaxies to the Milky Way (17 Mpc, Gavazzi et al. 1999; 
Mei et al. 2007), where multifrequency data covering the whole electromagnetic spectrum are now available down to the dwarf 
galaxy population (GUViCS, Boselli et al. 2011; NGVS, Ferrarese et al. 2012; HeViCS, Davies et al. 2010; ALFALFA, Giovanelli et al. 2005).
These works consistently indicate 
ram pressure as the dominant process responsible for the gas stripping and the quenching of the star formation activity of star forming systems
recently accreted by the cluster (Cayatte et al. 1990; Solanes et al. 2001; Gavazzi et al. 2013; Boselli et al. 2008ab, 2014b). 
The Virgo cluster has another major quality: thanks to its proximity the angular resolution of multifrequency data is 
comparable to that obtained in high-resolution simulations. The comparison of the kinematic and spectrophotometric
properties of several bright Virgo galaxies has been indeed crucial for the identification of the perturbing process 
(Vollmer et al. 1999, 2000, 2004, 2005, 2006, 2008a, 2008b, 2009, 2012; Vollmer 2003; Kenney et al. 2004; 
Boselli et al. 2005, 2006; Crowl \& Kenney 2008; Abramson et al. 2011; Kenney et al. 2014; Abramson \& Kenney 2014; Cortes et al. 2015).

A representative case is NGC 4569, the most massive late-type galaxy of the cluster ($M_*$ $\simeq$ 10$^{10.5}$ M$\odot$) located at $\sim$ 1.7 degree north-east from M87 
(corresponding to 0.32 virial radii $R_{vir}$ from the cluster core, see Fig. 
\ref{Virgo}). The study of the kinematic properties of the galaxy derived from HI data combined with simulations 
suggest that the galaxy has undergone a ram pressure stripping event with a peak of efficiency $\sim$ 300 Myr ago (Vollmer et al. 2004).
A similar result ($\sim$ 100 Myr) has been obtained by comparing the observed radial truncation of the different gaseous and stellar
components with multizone chemo-spectrophotometric models of galaxy evolution 
that are tailored to take into account the effects of ram pressure and starvation (Boselli et al. 2006; Crowl \& Kenney 2008).
This galaxy, however, as most of the massive galaxies in the nearby universe, is also characterised by a nuclear activity
with an associated outflow of gas (see sect. 6.2). It is thus an ideal candidate to study the relative contribution of 
ram pressure stripping and nuclear feedback to the stripping process of cluster galaxies.
For this purpose we have obtained a deep narrow band H$\alpha$+[NII] image of NGC 4569 and its surrounding regions with MegaCam at the CFHT,
that we combine here with an umpublished long-slit spectrum obtained at Calar Alto in 2001.
Deep H$\alpha$ imaging is used to search for long tails of ionised gas, the most direct witness of 
an ongoing ram pressure stripping event (e.g. Gavazzi et al. 2001). At the same time, the impact of feedback can be quantified by studying the properties of 
the ionised gas associated to the nuclear outflow.

\section{Observations}
   \begin{figure*}
   \centering
   \includegraphics[width=18cm]{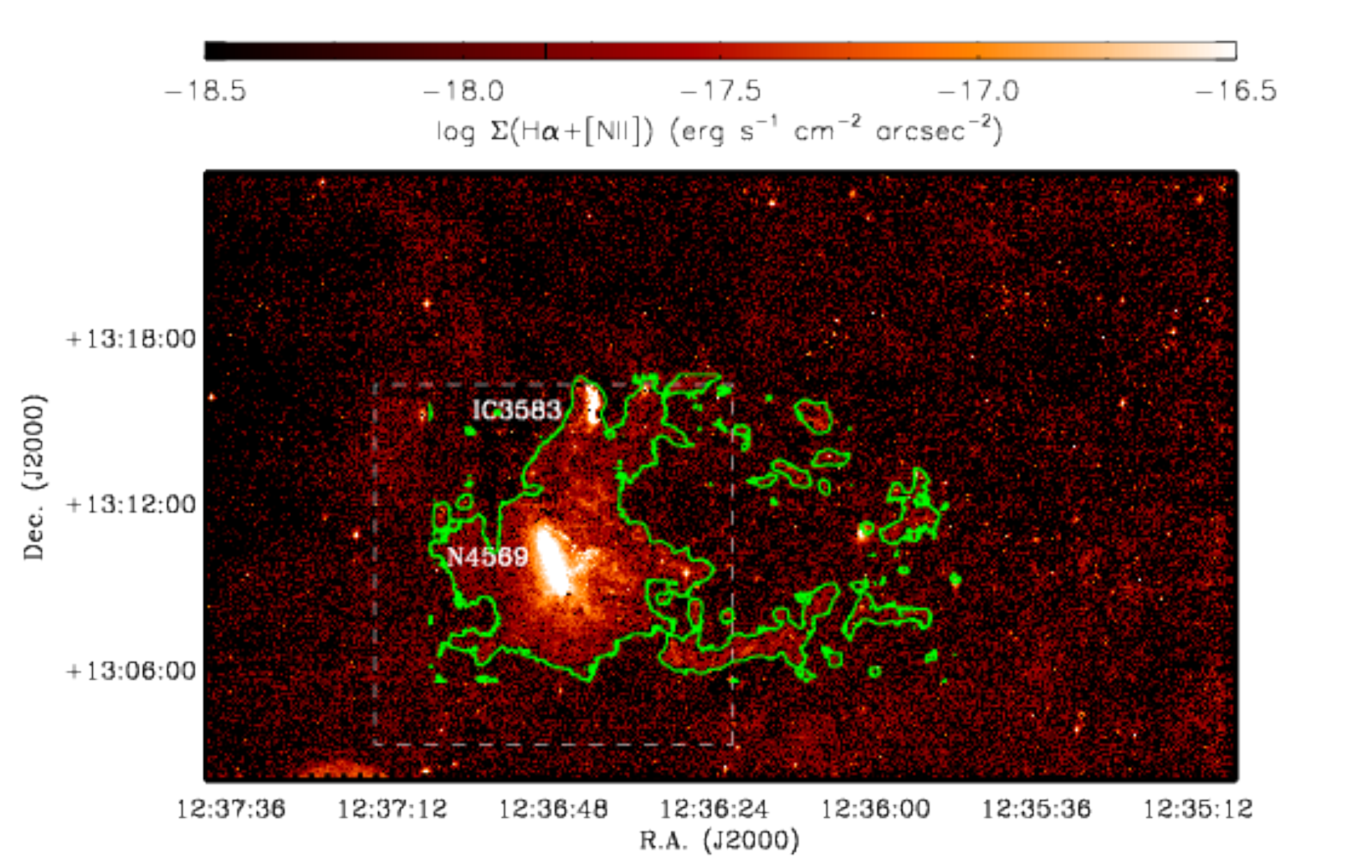}
   \caption{The CFHT MegaCam continuum-subtracted H$\alpha$+[NII] image of NGC 4569 and IC 3583 smoothed with a 5x5 median filter and masked from the emission of 
   foreground stars. Contours are drawn from the 10$^{-18}$ erg s$^{-1}$ cm$^{-2}$ arcsec$^{-2}$ 
   H$\alpha$+[NII] surface brightness level.  The white box marks the area observed with the 1.2m OHP telescope shown in Figure \ref{OHP}.
 }
   \label{Net}
   \end{figure*}

   \begin{figure}
   \centering
   \includegraphics[width=9cm]{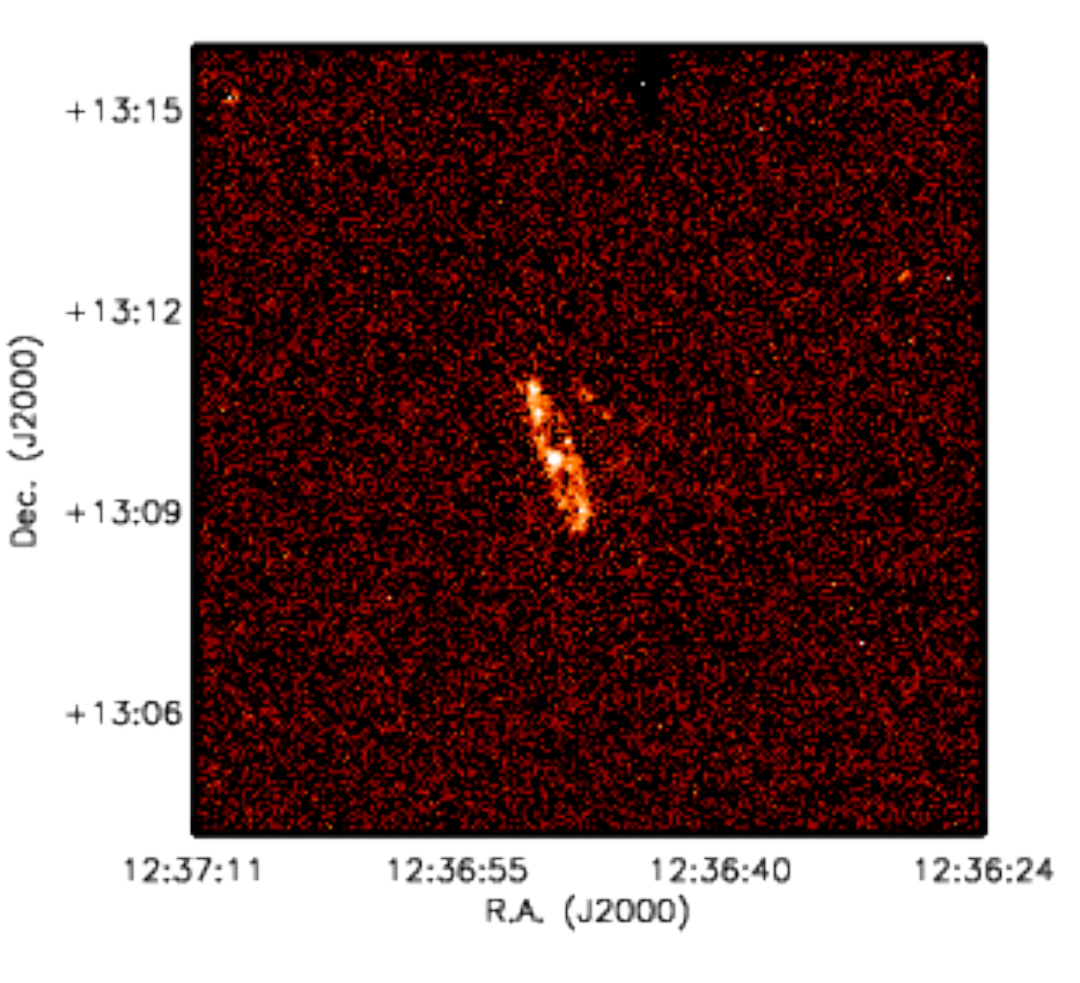}
   \caption{The H$\alpha$+[NII] image of NGC 4569 and IC 3583 obtained with a 30 min ON-band exposure at the 1.2m telescope of the Observatoire de Haute Provence, from Boselli \&
   Gavazzi (2002).  The image covers the area shown as the white box in Figure \ref{Net}.
 }
  \label{OHP}%
  \end{figure}

\subsection{Narrow band imaging}

The observations were carried out in May 2015 using MegaCam at the CFHT. NGC 4569, which has a recessional velocity of -221 km s$^{-1}$, was observed in the narrow band
filter MP9603 centered on the H$\alpha$ line ($\lambda$ = 6590 \AA; $\Delta \lambda$ = 104 \AA, $T$ = 93\%). The transmissivity $T$
of the filter (-1140 $<$ $vel$ $<$ 3600 km s$^{-1}$) perfectly covers the range in recessional velocity of the 
whole Virgo cluster region (Binggeli et al. 1993; Boselli et al. 2014a). Because of its width, the filter encompasses the two [NII] 
lines at $\lambda$ = 6548 and 6584 \AA. The stellar continuum was measured through the broad-band
$r$ filter. Since the purpose of the present observations was to detect low surface brightness features associated to the galaxy,
the observations were done using the pointing macro QSO LDP-CCD7 optimised for Elixir-LSB observing and processing mode (see sect. 3).
This macro makes 7 different pointings around the galaxy using a large dithering as generally done in near-infrared imaging of extended sources. 
The final fully co-added image covers the central 40'$\times$30' at maximum depth, 
while a larger area is mapped at lower sensitivity. This mode has been extensively used over the recent 
years and applies well to narrow-band imaging data with long exposures (medium sky background levels). Integration times were of 660 sec per pointing
in the ON-band image, and 66 sec in the OFF-band $r$-band filter, thus the resulting integration time on the stacked image is of 4620 sec in H$\alpha$+[NII]
and 462 sec in $r$. 
The photometric
calibration of the data, taken in photometric conditions, has been done in the $r$ band following the standard MegaCam procedures. In the narrow band filter 
the photometric calibration was secured with the observation of the spectrophotometric standards Feige 34 and Hz44 (Massey et al. 1988). The observation of these standard
stars gives consistent results within 2\%. The images were taken in good seeing conditions (0.80 arcsec in $r$ and 0.86 in H$\alpha$+[NII]).

 \begin{figure*}
   \centering
   \includegraphics[width=18cm]{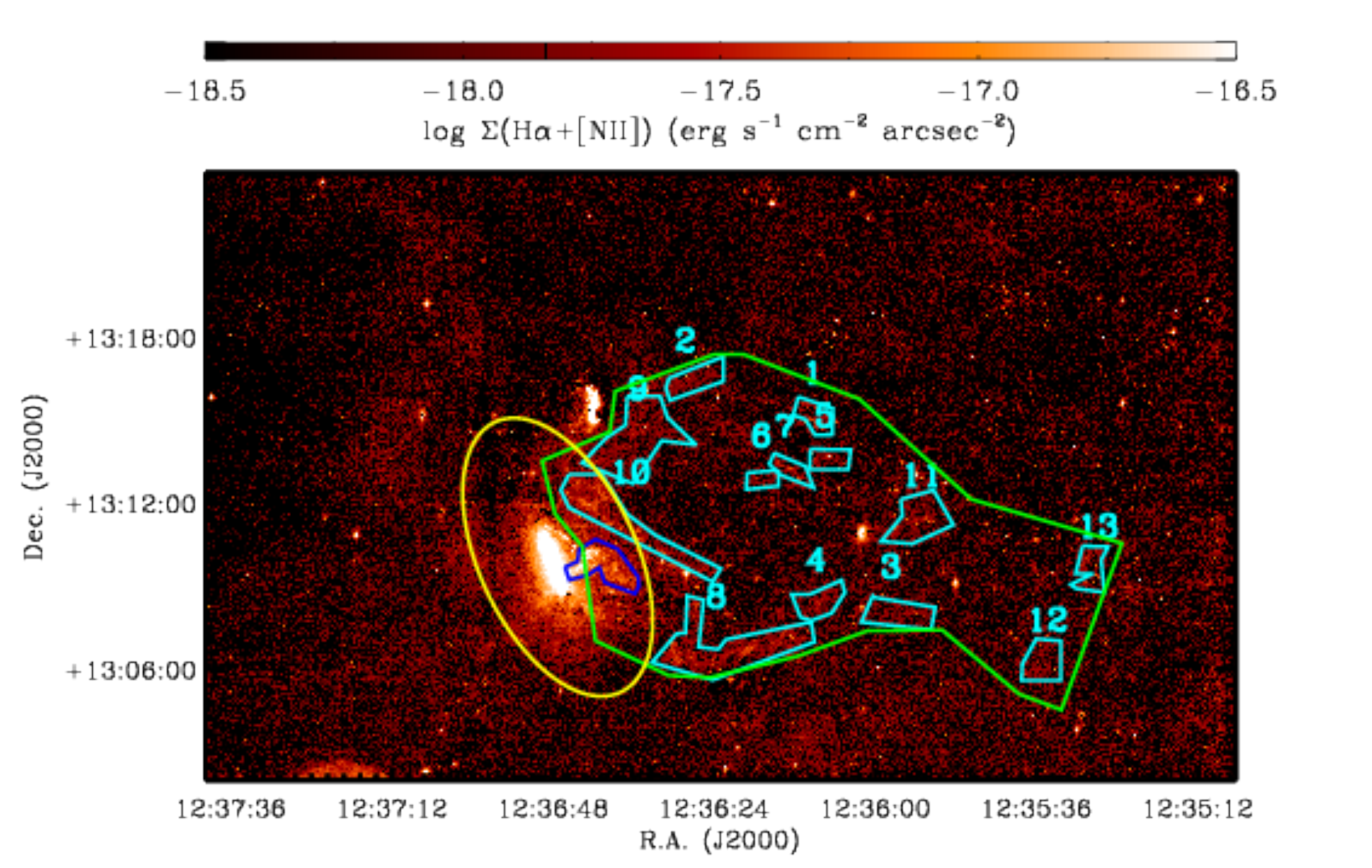}
   \caption{The CFHT MegaCam H$\alpha$+[NII] image of NGC 4569 and IC 3583 smoothed as in Fig 1. The different low surface brightness regions with H$\alpha$+[NII] emission are indicated with cyan polygons,
   and their surface brightness is listed in Table \ref{LSB}. The yellow elliptical aperture indicates the adopted extension of the galaxy within which its total H$\alpha$+[NII] flux has been 
   measured (Table \ref{gal}). The total emission within the tail has been measured within the large green polygon, while that of the nuclear outflow within the blue polygon.
 }
   \label{regions}%
   \end{figure*}

\subsection{Long slit spectroscopy}

NGC 4569 was observed in 2001 with the Calar Alto 3.5m telescope using the Twin spectrograph.
The galaxy was observed with the T05 and T06 gratings in the blue (4430 \AA) and red (6562 \AA) arm of the 
spectrograph with a dispersion of 36 \AA/mm  and a spatial sampling of 0.56 arcsec/pixel. 
Observations were obtained with a 30 min exposure using a 4 arcmin long slit of width 1.2 arcsec centered on the nucleus of the galaxy and 
oriented to a PA of 105$^o$ as indicated in Fig. \ref{Netcore}.
Due to the low signal to noise and sensitivity in the blue arm  
we were not able to obtain reliable measurements for a detailed
and satisfactory analysis of the data.

\section{Data reduction}

\subsection{Narrow band imaging}

The data have been reduced using the Elixir-LSB package (Cuillandre et al. in preparation), a pipeline expressly developed
to minimise the contribution of the scattered light component in MegaCam images. 
The efficiency of this observing strategy and data reduction procedures have been proven by the
detection of very low surface brightness features in the tidal tails associated to early-type galaxies in the MATLAS (Duc et al. 2011; 2015)
and in the NGVS surveys (Ferrarese et al. 2012; Mihos et al. 2015).\\

The best ON- and OFF-band images are then combined to produce a H$\alpha$+[NII] (continuum-subtracted) frame (Fig. \ref{Net}). 
The ON- and OFF-band frames are scaled using ad-hoc normalisation factors $n$ depending on the typical colour of the galaxy (Spector et al. 2012). 
The colour-dependent normalisation has been determined using 15 stars of different colour in the observed field. 
The photometric calibration of the image obtained using the spectrophotometric standard stars was checked 
using a few galaxies in the field observed during previous targeted narrow-band imaging observations or using nuclear spectroscopy from the SDSS (see Table \ref{gal}).
This comparison gives fairly consistent results (0.25 dex).

The pixel size in the raw images is of 0.187 arcsec/pixel. To increase the signal-to-noise we rebinned the images by a factor of three (0.561 arcsec/pixel),
and later smoothed them using a median 5x5 pixel filter. 
At low level counts the continuum-subtracted image also shows several low surface brightness features. Some of them are clearly associated to the reflection of bright stars in the
field, not fully removed by the Elixir-LSB data reduction procedure. Fortunately these are far from the target galaxy and only marginally affect the
present analysis. The image also shows filamentary structures in the west of NGC 4569 forming a double tail clearly associated to the galaxy (see next section).
The pixel 1 $\sigma$ rms of the resulting rebinned image is of 3.8 $\times$ 10$^{-18}$ erg cm$^{-2}$ sec$^{-1}$ arcsec$^{-2}$.
Because the signal is extended on scales of a few arcseconds in the extended filaments detected in H$\alpha$+[NII], once smoothed, the image is sufficiently deep to detect
features with a surface brightness of $\simeq$ 10$^{-18}$ erg cm$^{-2}$ sec$^{-1}$ arcsec$^{-2}$ (Fig. \ref{Net}). The comparison of the continuum-subtracted image of the galaxy (Fig. \ref{Net})
with the one obtained in a 30 min exposure with the 1.2 meter OHP telescope (Fig. \ref{OHP}, Boselli \& Gavazzi 2002)  
underlines the exquisite quality of the present image.
The total H$\alpha$+[NII] flux of the stellar disc measured within an elliptical aperture arbitrary defined to minimise the contribution of IC 3583 (Fig. \ref{regions})
is log$f(H\alpha +[NII])$ = -11.85$\pm$0.03 erg s$^{-1}$ cm$^{-2}$, which is consistent with previous estimates (see Table \ref{gal}).\\

\subsection{Long slit spectroscopy}

Reduction was performed using the {\it longslit} package 
in IRAF. We used dome flats to create the spectroscopic flat field.
Wavelength calibration was performed using the Thorium-Argon lamp. 
The tilt of the dispersion axis with respect to the CCD 
rows was corrected using a template created from observations of standard 
stars and the galaxy core at different positions along the slit. 

\begin{figure}
   \centering
   \includegraphics[scale=0.63]{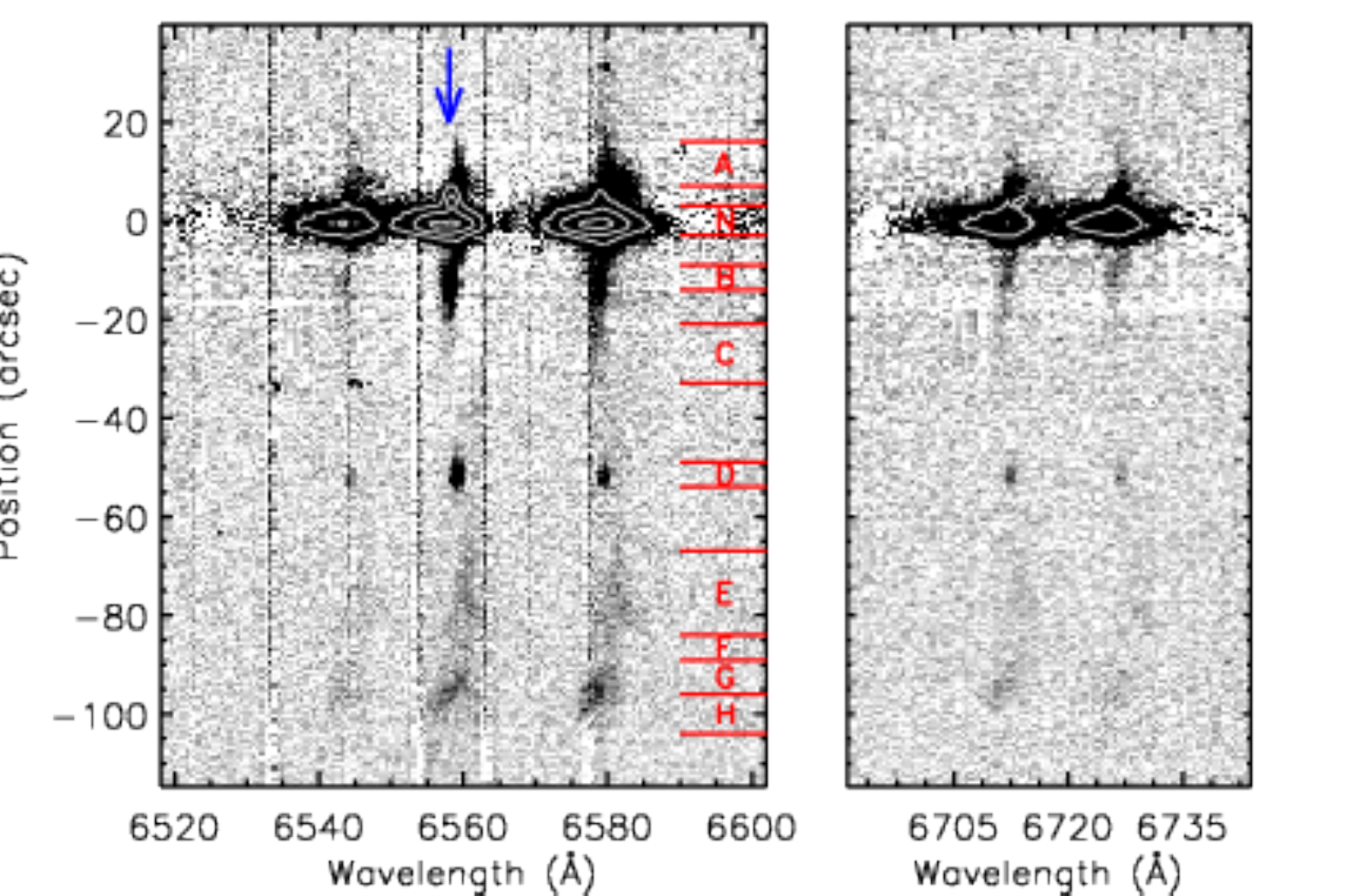}
   \caption{The 2D continuum subtracted spectrum of NGC 4569 zoomed on the H$\alpha$ and [NII] lines (left) and on the [SII] doublet (right) 
   obtained at Calar Alto with a slit of width 1.2 arcsec positioned along the minor axis on the spur of ionised gas
   as indicated in Fig. \ref{Netcore}. The red labels shown on the left panel correspond to those shown in Fig. \ref{Netcore} and 
   indicate the range of pixels used to extract 1D spectra to derive the 
   physical parameters given in Table \ref{tab:ratios}. The blue vertical arrow shows the mean recessional velocity of the galaxy (-221 km s$^{-1}$)
   derived from HI data.
 }
   \label{2D}
   \end{figure}

From the sky subtracted, continuum subtracted, and wavelength calibrated 2D spectrum (shown in Figure \ref{2D}) we extract 
1D spectra in the regions given in Table \ref{tab:ratios} using median statistics. Variance spectra 
are extracted in the same positions along the slit by assuming Poisson statistics on the 
un-skysubtracted 2D spectrum. The spectral resolution ($R=\lambda/\Delta \lambda$) is obtained by 
fitting 18 bright isolated sky lines in the spectrum in a position close to the galaxy nucleus. 
We then fitted a 3rd order polinomial to those points to obtain a model of $R$ vs. wavelength which
is then used in the fitting procedure to correct the width of the lines for the instrumental resolution.
The spectral resolution of the data is $R=6290$ (FWHM = $46~\rm{km~s^{-1}}$ or 1.02 $\AA$) at H$\alpha$. 

We fit the emission line of the spectrum with the 
{\sc kubeviz} software\footnote{http://www.mpe.mpg.de/\~{}mfossati/kubeviz/} (Fossati et al. 2015). 
This code uses ``linesets'', defined as groups of lines that are fitted 
simultaneously. Each lineset is described by a combination of 1D Gaussian functions 
where the relative velocity separation of the lines is kept fixed. In this work we fit two linesets, the first
made of H$\alpha$ and [NII] $\lambda \lambda 6548,6584$ and the second made of [SII] $\lambda \lambda 6716,6731$. 
Furthermore, the flux ratio of the two [NII] lines is kept constant in the fit to the ratios in Storey \& Zeippen (2000). 
The continuum level is evaluated during the fit procedure with an initial guess estimated in two symmetric 
windows around each lineset. 
During the fit, {\sc kubeviz} takes into account the noise from the variance spectra, thus optimally suppressing 
sky line residuals (which bracket the H$\alpha$ line at the redshift of NGC4569). However, the adopted 
variance underestimates the real error, most notably because it does not account for correlated noise introduced 
in the reduction procedure and extraction of 1D spectra. We therefore renormalise the final errors 
on the line fluxes and kinematical parameters assuming a $\chi^2 =1$ per degree of freedom.
The resulting emission line measurements are given in Table \ref{tab:ratios}. We verified that the kinematical
parameters are consistent within the uncertainties if we compare the two linesets (H$\alpha$+[NII] and [SII]). None 
the less the values listed in Table \ref{tab:ratios} are obtained from the first lineset because the 
uncertainties are smaller due to the brighter lines. Diagnostic line ratios are also given whenever the 
signal-to-noise of all the lines involved in the ratio is greater than 3.

   \begin{figure*}
   \centering
   \includegraphics[width=19cm]{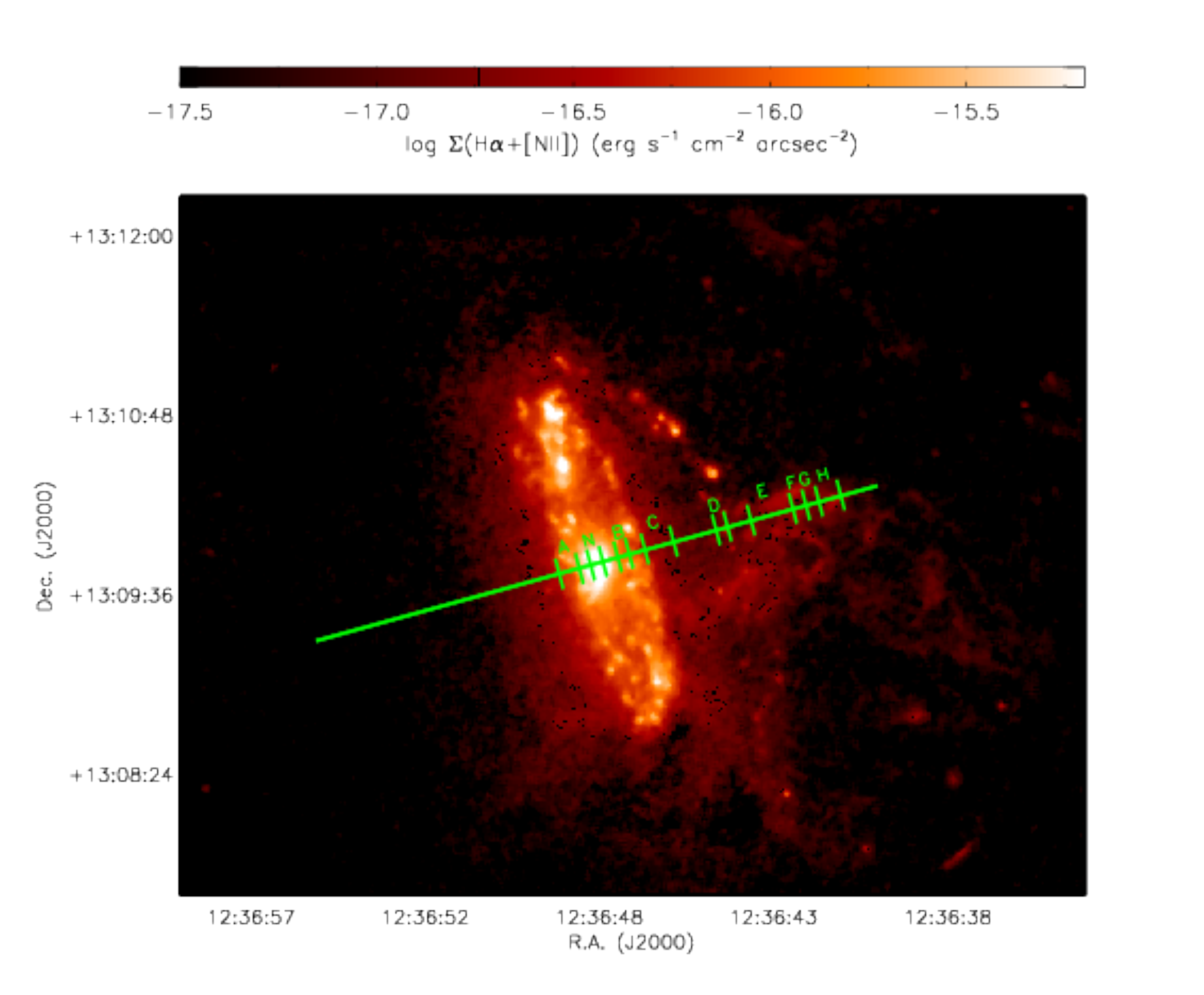}
   \caption{The CFHT MegaCam H$\alpha$+[NII] image of the star forming disc and of the nuclear outflow of NGC 4569. The green line indicates the position of the long slit 
   in the Calar Alto spectroscopic observations. The different letters indicates the various regions used to extract 1D-spectra and derive the physical parameters listed in Table
   \ref{tab:ratios}. They corresponds to those indicated in Fig. \ref{2D}.
 }
   \label{Netcore}
   \end{figure*}

\section{Physical parameters}

\subsection{Narrow band imaging}

The deep H$\alpha$+[NII] image of NGC 4569 shows several filamentary features extending from the disc of the galaxy to $\sim$ 80 kpc projected distance 
to the west. These filaments are dominated by two main features as often observed in cluster galaxies with extended tails of ionised gas (Sun et al. 2007,
2010; Yagi et al. 2010). 
The H$\alpha$+[NII] fluxes of NGC 4569 and of its associated low surface brightness extended features (Table \ref{LSB}), 
as well as that of the other galaxies in the frame (Table \ref{gal}),
are determined by measuring the counts in the ON- and OFF-band frames (after removing the contribution of unwanted foreground stars in the field). 
The flux of all these targets has been extracted using the QPHOT IRAF task whenever possible. For sources with asymmetric distributions or with evident nearby
companions which might significantly contaminate the flux we used  
the FUNTOOLS analysis package on DS9. For these sources we manually defined elliptical apertures adjusted to fit the full light profile of each target galaxy
and to select uncontaminated regions for the determination of the sky background. Using polygons, we identified in the image smoothed with a median 
filter of 5$\times$5 pixels a number of low surface brightness features labeled in Fig. \ref{regions}. Their flux has been extracted using FUNTOOLS, while the local sky
background was estimated in adjacent polygons. We also extracted the total flux of all the extended filaments using a large polygon covering the western part of the galaxy,
with its associated sky background measured in several circular apertures located in uncontaminated regions around it. The uncertainty on these measurements has been
determined using the prescription given in Boselli et al. (2003) which is optimised for extended sources where the uncertainty is dominated by large scale fluctuations of
the sky background (see also Ciesla et al. 2012). The properties of these regions are listed in Table \ref{LSB}.

The H$\alpha$ luminosity of the different galaxies and of the extended features can be determined once the observed fluxes are corrected for [NII] contamination. 
This is relatively easy for the emission over the discs of the two galaxies NGC 4569 and IC 3583, an irregular galaxy at a projected distance of $\sim$ 6 arcmin, 
for which integrated spectroscopy is available from Gavazzi et al. (2004) and Boselli et al.
(2013)\footnote{We use for this purpose the updated value of [NII]/H$\alpha$ = 0.97 for NGC 4569 given in Boselli et al. (2015).}. 
The physical conditions of the stripped interstellar medium might be significantly different than over the disc
of the galaxy, thus this ratio might significantly change. In particular, the radiation of the hot gas of the intracluster medium is expected to increase 
the [NII]/H$\alpha$ ratio. Recent spectroscopic observations done with the Multi Unit Spectroscopic Explorer (MUSE) of ESO 137-001, 
a galaxy in the Norma cluster undergoing a stripping process similar to the one occurring in NGC 4569, suggest that the [NII]/H$\alpha$ 
ratio in the tail is not typical of a photoionisation region ([NII]/H$\alpha$ $\simeq$ 0.3), but is slightly higher likely because the gas is partly excited by shocks or
heat conduction ([NII]/H$\alpha$ $\simeq$ 0.5, Fossati et al. 2015). 
Similar results have been derived from the spectroscopic observations of the tails of other cluster galaxies (Yagi et al. 2007; Yoshida et al. 2012; Merluzzi et al. 2013).
We thus assume [NII]/H$\alpha$ = 0.5 in the tails.

An estimate of the density of the ionised gas can be derived using the relation:

\begin{equation}
{L(H\alpha) = n_e n_p \alpha^{eff}_{H\alpha} V f h \nu_{H\alpha}}
\end{equation}

\noindent
(Osterbrock \& Ferland 2006) where $n_e$ and $n_p$ are the number density of electrons and protons, $\alpha^{eff}_{H\alpha}$
is the H$\alpha$ effective recombination coefficient, $V$ is the volume of the emitting region, $f$ the filling factor, 
$h$ the Planck's constant, and $\nu_{H\alpha}$ the frequency of the H$\alpha$ transition. The two variables $V$ and $f$ 
can be only crudely estimated from observations or from simulations. 

The stripped material is assumed to be 
distributed in a cylinder of diameter ($\simeq$ 50 kpc) and of height comparable to the 
extension of the observed tail of ionised gas. In the case of NGC 4569, the tail extends up to $\simeq$ 80 kpc on the plane of the sky.
Since the galaxy is blue-shifted ($vel$ = -221 km s$^{-1}$), we expect that it is crossing the cluster from the backside. It is  
conceivable that the observed tail of ionised gas is just a projection on the plane of the sky, thus that 80 kpc is a lower limit to the 
real height of the cylinder. The comparison of multifrequency observations of the galaxy with tuned models of gas stripping suggest that 
the galaxy underwent the peak of the stripping process $\simeq$ 100 Myr ago (Boselli et al. 2006). At a radial velocity of $\simeq$ 1100 km s$^{-1}$ 
with respect to the cluster centre\footnote{The mean velocity of cluster A, the Virgo substructure to which NGC 4569 belongs, is 
$vel$ $=$ 955 km s$^{-1}$ (Boselli et al. 2014a).}, the galaxy would have traveled $\sim$ 120 kpc along the line of sight. Summing these values quadratically, 
we expect that the physical extension of the stripped gas is $\sim$ 145 kpc, corresponding to $\sim$ 5.5 times the optical radius of the galaxy. 

The filling factor $f$ is another unconstrained parameter. 
In all galaxies where tails have been observed, the ionised gas has a structured distribution, with high density clumps of condensed material
often associated to star forming regions, otherwise extended in filamentary structures (Yagi et al. 2007, 2010, 2013). Similar structures are also present in
hydrodynamic simulations (e.g. Tonnesen \& Bryan 2010). In particular, the filamentary structures with double tails are reproduced in simulations 
whenever magnetic fields are taken into account (Ruszkowski et al. 2014, Tonnesen \& Stone 2014). 
It is thus most likely that $f$ $<$ 1, but its exact value is highly uncertain.
Consistently with previous works, which generally take 0.05 $<$ $f$ $<$ 0.1, we assume $f$ = 0.1. 

If we assume that the gas is fully
ionised, thus that $n_e$ = $n_p$, and $\alpha^{eff}_{H\alpha}$ = 1.17 $\times$ 10$^{-13}$ cm$^3$ s$^{-1}$ (Osterbrock \& Ferland 2006)
the density of the gas can be derived from eq. (1):

\begin{equation}
{n_e = \sqrt{\frac{L(H\alpha)}{\alpha^{eff}_{H\alpha}Vfh\nu_{H\alpha}}}}
\end{equation}

\noindent
Under these assumptions, the mean density of the ionised gas is $n_e$ $\simeq$ 5 $\times$ 10$^{-3}$ cm$^{-3}$, and the total mass
of the ionised gas is $M(_{tail}(H\alpha)$ $\simeq$ 3.2 $\times$ 10$^{9}$ M$_{\odot}$. Given the large uncertainty on the geometry of the gas, this is a
very rough estimate. We can also calculate the typical density of the gas within the different filaments labeled in Fig. \ref{regions}.
Assuming a cylindrical geometry, the typical density within the different filaments ranges between 2 and 4 $\times$ 10$^{-2}$ cm$^{-3}$,
while the total mass in the 13 identified regions is $\simeq$ 3.1 $\times$ 10$^{8}$ M$_{\odot}$\footnote{This value should be taken as a lower limit to the total mass of the ionised gas 
given that the 13 regions indicated in Fig. \ref{regions} do not include all the low surface brightness filaments associated to the galaxy.}.

\subsection{Long slit spectroscopy}

\subsubsection{Kinematics}

The spectral resolution of the Calar Alto observations is sufficient to study the kinematics of the ionised gas along the slit (see Table \ref{tab:ratios}). 
The 2D-spectrum of the galaxy (Fig. \ref{2D})  
shows an offset in velocity in the SE (region A, extending up to $\simeq$ 20 arcsec) and NW (region B, up to $\simeq$ 30 arcsec). 
This offset, which is observed also in the H$\alpha$ Fabry-Perot data of Chemin et al. (2006), is due to the rotation of the galaxy. This region corresponds
to the H$\alpha$ main body of the galaxy in Fig. \ref{Netcore}. The nuclear spectrum also shows a large velocity dispersion (132 km s$^{-1}$)
due to the turbulence in the gas probably induced by a nuclear activity (Ho et al. 1997). 
A high surface brightness region is detected at $\sim$ 50 arcsec from the nucleus in the NW direction (region D). This region, which is located in the projection of the
extended NW spiral arm, is present also in the Fabry-Perot data. The velocity dispersion of this region is low ($\simeq$ 10 km s$^{-1}$) and is typical of an 
HII region. Further out, emission is detected along the filaments of ionised gas at high velocity (up to $\simeq$ 130 km s$^{-1}$ in region E)
with respect to the nucleus. The kinematic of the gas, as shown in Fig. \ref{2D}, is totally disconnected from the rotation of the disc.
Further out (regions F, G), the recessional velocity decreases smoothly, then rapidly in region H reaching negative values. 
The velocity dispersion in these outer and diffuse regions is relatively high (70-90 km s$^{-1}$), indicative of the presence of turbulent motions.

\subsection{Line ratios}

We can also estimate how the [NII], H$\alpha$, and [SII] line ratios change along the slit in the different positions listed in Table \ref{tab:ratios}.
The [NII]/H$\alpha$ line ratio in the nucleus is 1.29$\pm$0.04, while [SII]/H$\alpha$ = 0.67$\pm$0.02. These values are typical of LINER galaxies (e.g. Ho et al. 1997).
The values of region D ([NII]/H$\alpha$ = 0.37$\pm$0.03; [SII]/H$\alpha$ = 0.26$\pm$0.02) are typical of HII regions, confirming that this high surface 
brightness spot is a star forming regions associated to the western spiral arm. The values of [NII]/H$\alpha$ and [SII]/H$\alpha$ measured in all other regions,
on the contrary, are in the range 0.7 $\lesssim$ [NII]/H$\alpha$  $\lesssim$ 1.4 and 0.4 $\lesssim$ [SII]/H$\alpha$  $\lesssim$ 0.9. Although uncertain,
these values are too high to be produced by stellar photoionisation only, but rather require the contribution of an hard radiation field as the one produced by a nearby nucleus
or by slow shocks (Tuelmann et al. 2000; Allen et al. 2008; Rich et al. 2011). In the spur of ionised gas the [SII]$\lambda$6716/[SII]$\lambda$6731 
line ratio is $\simeq$ 1.4 suggesting that the electron density of gas is low ($n_e$ $\lesssim$ 10 cm$^{-3}$). The exact value is poorly constrained given the large 
uncertainty on the [SII] ratio and the saturation in the intensity ratio vs. electron density relation (Osterbrock \& Ferland 2006).

   \begin{figure*}
   \centering
   \includegraphics[width=18cm]{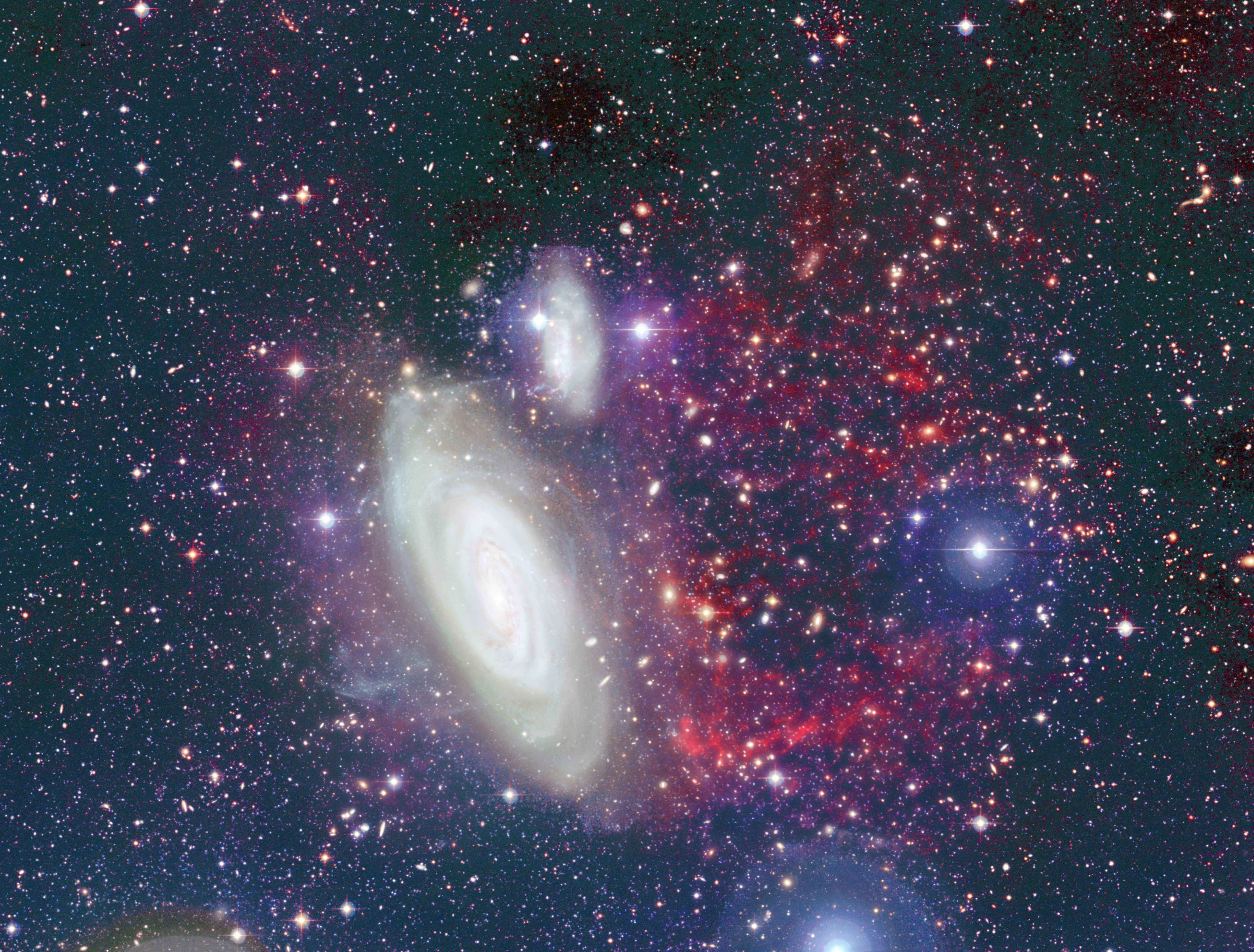}
   \caption{The pseudo-colour image of NGC 4569 and IC 3583 obtained combing the CFHT MegaCam NGVS optical $g$ (blue) and $i$ (green)
   images with the H$\alpha$+[NII] narrow band image (red). North is up, east left.
 }
   \label{RGB}%
   \end{figure*}

\section{Comparison with multifrequency data}

\begin{figure*}
   \centering
   \includegraphics[scale=0.9, angle=90]{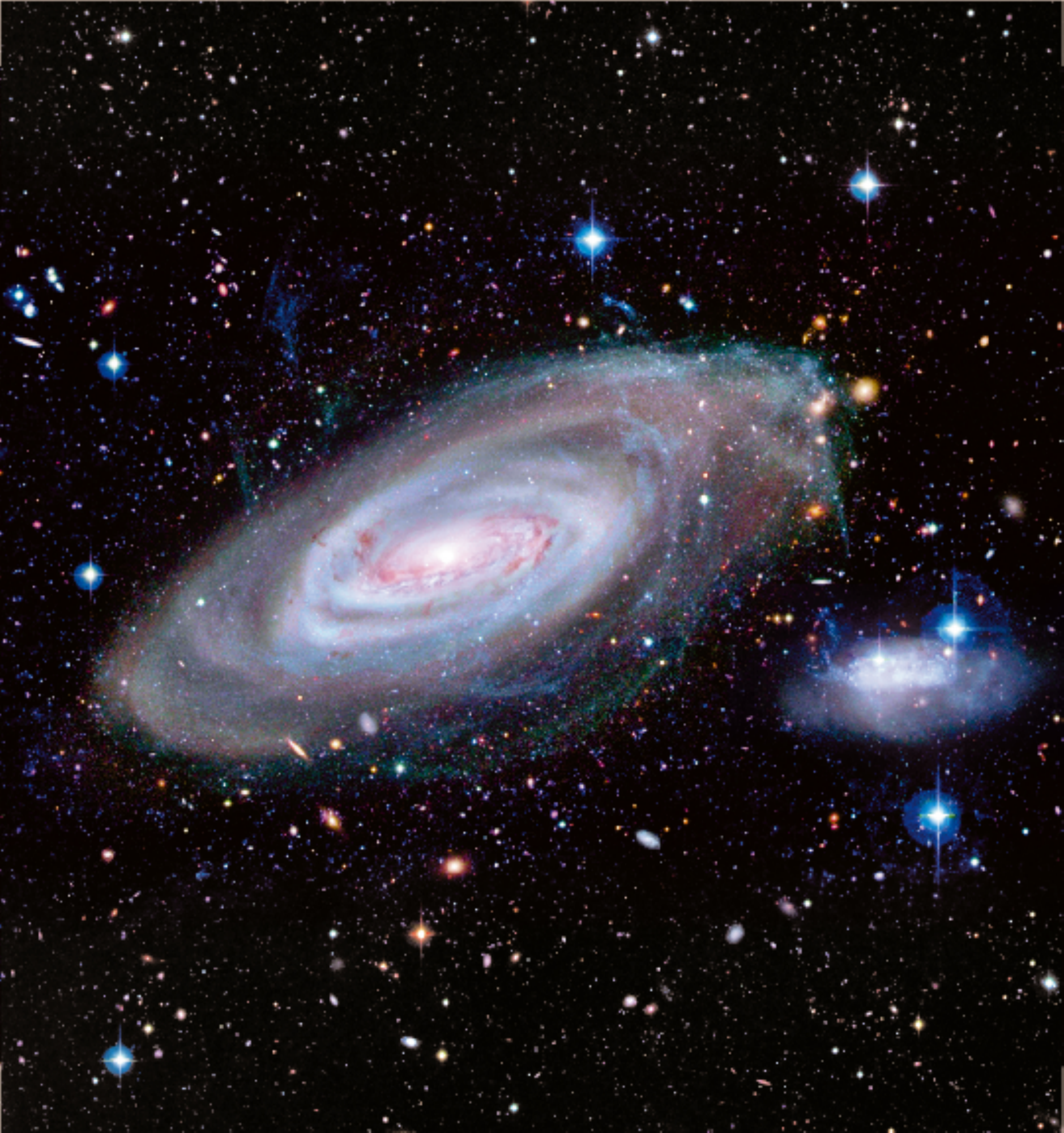}
   \caption{The pseudo-colour image of NGC 4569 and IC 3583 obtained using the CFHT MegaCam NGVS (Ferrarese et al. 2012) optical $u$ (blue), $g$ (green), and $i$ (red)
   images. North is up, east left.
 }
   \label{NGVS}%
   \end{figure*}

\begin{figure*}
   \centering
   \includegraphics[width=18cm]{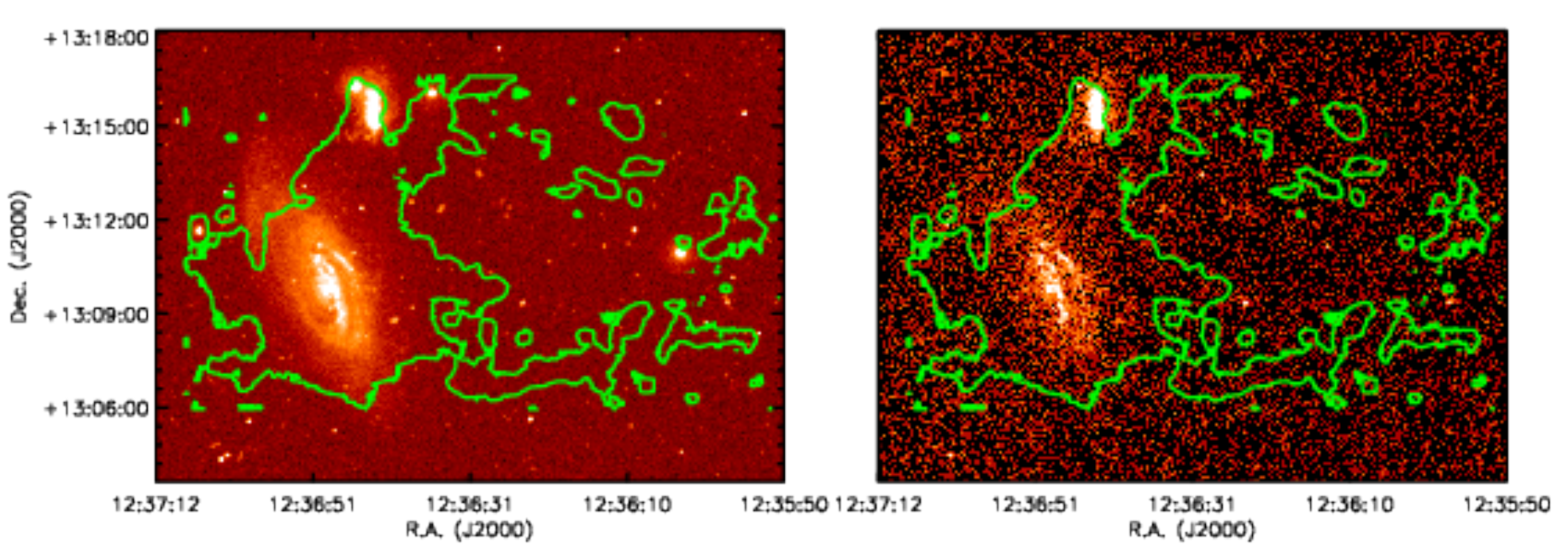}
   \caption{The GALEX NUV (upper) and FUV (lower) images of NGC 4569 with contour levels showing the H$\alpha$+[NII] surface brightness level of 10$^{-18}$ erg s$^{-1}$
   cm$^{-2}$ arcsec$^{-2}$. 
    }
   \label{UV}%
   \end{figure*}

\begin{figure}
   \centering
   \includegraphics[width=9cm]{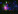}
   \caption{The pseudo-colour image of NGC 4569 and IC 3583 obtained combaining XMM 0.4-1.3 keV X-rays (blue), GALEX NUV (green), and CFHT MegaCam H$\alpha$+[NII] (red)
   images. North is up, east left.
 }
   \label{comp}%
   \end{figure}

High-quality multifrequency data covering the whole electromagnetic spectrum, from X-ray to radio, of NGC 4569 are available
in the literature. These data are crucial for a comparison with the H$\alpha$ data obtained in this work to identify
the perturbing mechanism affecting the galaxy. Excellent quality $ugiz$ imaging data of NGC 4569 have been obtained as part of the NGVS survey 
using MegaCam at the CFHT (Ferrarese et al. 2012). The pseudo-colour optical image of the galaxy is shown in Fig. \ref{RGB} and \ref{NGVS}.
The optical image does not show any diffuse stellar emission down to a surface brightness limit of $\simeq$ 29 $\mu_g$ mag arcsec$^{-2}$ (AB system) 
associated to the stripped ionised gas located in the west of the galaxy. The lack of low surface brightness stellar features, which are generally formed
during gravitational interactions with nearby companions, suggests that the gaseous component is removed from the galaxy through the interaction with 
the hot and dense intracluster medium. The comparison of the H$\alpha$ frame with the optical NGVS (Fig. \ref{RGB}) and the FUV and NUV (Figs. \ref{UV} and \ref{comp}) images of the galaxy 
obtained as part of the the GUViCS survey of the cluster (Boselli et al. 2011) also indicates that the ionised gas is only diffuse and does not have any clumpy structure
suggesting the presence of extraplanar HII regions. The smallest resolved features in the H$\alpha$+[NII] image 
have a filametary structure of thickness larger than 200 pc. We remark that the NUV image of the galaxy and its surrounding regions has been obtained with a 
very long exposure (16993 sec) and it is thus very sensitive to extraplanar star forming regions such as those observed around M49 (Arrigoni-Battaia et al.
2012) or VCC 1217 (Hester et al. 2010; Fumagalli et al. 2011; Kenney et al. 2014).  The H$\alpha$+[NII] surface brightness of the extraplanar HII regions associated to these two galaxies is $\sim$ 
2 $\times$ 10$^{-17}$ erg cm$^{-2}$ sec$^{-1}$ arcsec$^{-2}$, thus well above the detection limit of our image. 
The limiting sensitivity in the NUV image is of $\simeq$ 29 AB mag arcsec$^{-2}$, thus deep enough to detect low surface brightness features such as those
observed in the tidal tails of NGC 4438 (Boselli et al. 2005).\\

\begin{figure*}
   \centering
   \includegraphics[width=18cm]{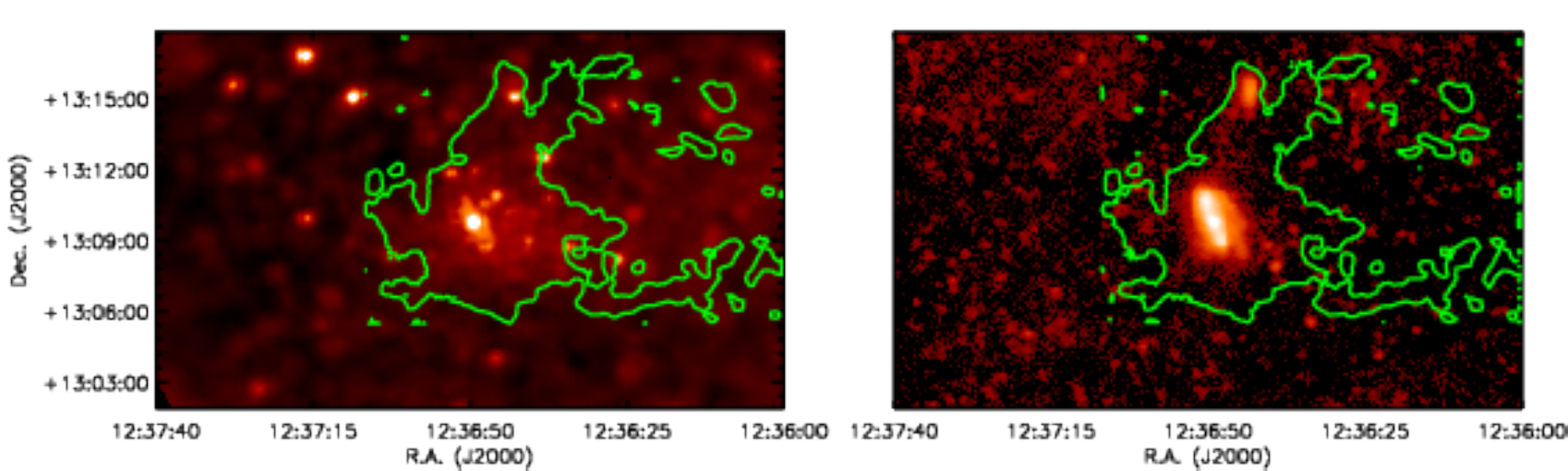}
   \caption{The 0.4-1.3 keV background subtracted and exposure corrected X-ray (left) XMM image and the far-infrared 250 $\mu$m (right; Ciesla et al. 2012) images of NGC 4569 with contour 
   levels showing the H$\alpha$+[NII] surface brightness level of 10$^{-18}$ erg s$^{-1}$ cm$^{-2}$ arcsec$^{-2}$. 
 }
   \label{XFIR}%
   \end{figure*}

The distribution of the atomic gas has been mapped using the VLA in C short configuration by Chung et al. (2009). The HI image,
which has an angular resolution of 15-16 arcsec and a column density sensitivity of 3-5 $\times$ 10$^{19}$ cm$^{-2}$ (3 $\sigma$ per channel), 
shows that the HI gas is located within the stellar disc of the galaxy and has a 
truncated radial distribution typical of HI-deficient cluster galaxies (Cayatte et al. 1994; Chung et al. 2009).  This is also the case for the distribution of the 
molecular gas (Helfer et al. 2003) and of the hot and cold dust components as derived from WISE, \textit{Spitzer}, and \textit{Herschel}\footnote{NGC 4569 has been observed during the
\textit{Herschel} Reference Survey, Boselli et al. (2010). Most of these multifrequency data are available on a dedicated database: http://hedam.lam.fr/HRS/} images 
(Boselli et al. 2006; Cortese et al. 2014; Ciesla et al. 2012; Fig. \ref{XFIR}). The atomic gas and the dust component are not detected 
along the tails of ionised gas.


Rosat, \textit{XMM-Newton}, \textit{Chandra} X-ray and VLA 1.4, 4.86 GHz radio continuum images of NGC 4569 show extended features perpendicular 
to the stellar disc of 
the galaxy (Tschoeke et al. 2001; Chyzy et al. 2006; Grier et al. 2011; Wezgowiec et al. 2011, 2012). The extension of the hot gas traced by the 0.2-1 keV \textit{XMM-Newton} X-ray image, 
however, is smaller than that of the ionised gas (Fig. \ref{XFIR} and \ref{comp}). The 20 cm radio continuum emission, instead, is 
limited to the inner 22 kpc and does not morphologically match the ionised gas. 

\section{Discussion}

\subsection{The galaxy}

NGC 4569 shows a truncated star 
forming disc with respect to the distribution of the old stellar populations (Boselli et al. 2006),  
with a prominent spiral arm starting from the north of the galaxy and extending in the west direction. 
The excellent quality of the CFHT image shows resolved HII regions along the disc and the western spiral arm.
A concentration of HII regions is present on the north-east and south-west star forming disc, suggesting
a grand design spiral pattern drawn by two major spiral arms. Independent tracers of star formation, including the H$\alpha$ flux derived in this work, 
can be converted under some assumptions into star formation rates (e.g. Kennicutt 1998, Boselli et al. 2009). The star formation rate 
measured over the whole disc of NGC 4569 is $SFR$ $\simeq$ 2 M$_{\odot}$ yr$^{-1}$  assuming a Salpeter IMF (Boselli et al. 2015).

The wavelength dependent truncation of the stellar disc and of the gas (atomic and molecular) and dust components observed in NGC 4569 has 
been explained as due to a recent ($\simeq$ 100 Myr) ram pressure stripping event able to radially remove the gaseous component and gradually quench
the activity of star formation of the galaxy in the outer regions (Boselli et al. 2006). A similar timescale ($\simeq$ 300 Myr) has been derived
from the study of the kinematic properties of the atomic gas (Vollmer et al. 2004) and from the analysis of 2-D optical spectra (Crowl \& Kenney 2008). 
The physical and kinematic properties of the main body of NGC 4569 thus consistently indicate that the galaxy underwent 
a recent ram pressure stripping event. We want to see whether this evolutionary picture can also explain the presence of the low surface brightness 
features detected in the H$\alpha$+[NII] narrow band image described in the previous section.

\subsection{The nuclear outflow}

The deep MegaCam H$\alpha$+[NII] image shows a diffuse and extended halo of ionised gas around the galaxy. It also shows a prominent 
plume in the western side perpendicular to the disc. This plume is located south of the minor axis in the region between the 
star forming disc and the prominent western spiral arm, and along the minor axis outside it, identified with a blue polygon in Fig. \ref{regions}. 
Its total extension is $\simeq$ 100 arcsec (8 kpc), and its total flux is 
log$f(H\alpha +[NII])$ = -13.07 erg s$^{-1}$ cm$^{-2}$ (see Table \ref{LSB}). 
The presence of a prominent dust lane in the western side of the galaxy (Fig. \ref{NGVS}), hidden by the bulge in the eastern side, 
suggests that the western side is the near side. The spur of gas observed in the western side has a higher recessional velocity with respect to the nucleus of the galaxy
and is thus an outflow. The nuclear outflow is probably powered by a nuclear starburst
(Barth et al. 1998; Maoz et al. 1998; Barth \& Shields 2000, Tschoeke et al. 2001, Chyzy et al. 2006).
The spectral synthesis analysis of the nuclear HST spectra carried out by Barth \& Shields (2000), who classified it as a typical LINER/HII transition nucleus, 
dated the nuclear ($\lesssim$ 30 pc) starburst to $\sim$ 3-6 Myr (Barth \& Shields 2000,
Gabel \& Bruhweiler 2002). The presence of A-type supergiants in the inner $\sim$ 300 pc suggests that a second starburst occurred more than 15 Myr ago (Keel 1996).
A contribution from an AGN cannot be fully excluded given the presence of a compact source in the soft \textit{Chandra} X-ray image of NGC 4569 
(Grier et  al. 2011). 
A dominant AGN activity, however, is ruled out by the lack of a point-like nuclear source in the radio
continuum (Hummel et al. 1987; Neff \& Hutchings 1992) and in the ROSAT X-ray hard band image of the galaxy (Tschoeke et al. 2001). 


If we assume that the gas is mainly photoionised we can use eq. 2 to estimate the total density and mass of the ionised gas 
in the outflow. Assuming that the gas in the outflow is in a cylinder of 60 arcsec (5 kpc) diameter and 96 arcsec (8 kpc) projected length, [NII]/H$\alpha$ $\simeq$ 1, and a filling factor $f$ = 0.1,
the density of the ionised gas is $n_e$ $\simeq$ 9.7 $\times$ 10$^{-2}$ cm$^{-3}$ and its total mass $M_{out}(H\alpha)$ $\simeq$ 3.4 $\times$ 10$^7$ M$_{\odot}$ (see Table \ref{masse}). 
This density is consistent with the mean electron density $n_e$ $\lesssim$ 10 cm$^{-3}$ derived by the poorly constraining [SII]$\lambda$6716/[SII]$\lambda$6731 
line ratio of the outflow ($\simeq$ 1.4). At this density,
the recombination time is $t_{rec}$ $\simeq$ 1 Myr, shorter than the age of the nuclear starburst. For an outflow velocity of $\simeq$ 130 km s$^{-1}$, 
the ejected gas should have traveled only a distance of $\simeq$ 0.65 kpc in 5 Myr, and $\gtrsim$ 2 kpc if powered by an older ($>$ 15 Myr) 
starburst episode that is necessary to explain the presence of A-type supergiants in the inner 300 pc (Keel 1996). These distances are too small compared to the typical extension of
the spur of ionised gas ($\simeq$ 8 kpc). It is thus conceivable that the gas in the outflow has been ionised by other mechanisms than photoionisation from the
central starburst. These can be identified as the nuclear activity or the shock induced by the turbulence in the outflow.\\

We can also make a rough estimate of the total (potential and kinetic) energy of the outflow. Considering the outflow as a cylinder of physical length $H$ and radius $r$, 
its potential energy can be derived using the relation:

\begin{eqnarray}
{d\Phi_{out} = G \frac {M_{N4569+out}}{h}dM_{out} =}\nonumber\\
{= G (\rho_{N4569} + \rho_{out}) (\pi r^2)^2 H \rho_{out}\frac{dh}{h}}
\end{eqnarray}
 
\noindent 
where $\rho_{N4569}$ and $\rho_{out}$ are the mean mass density of the galaxy and of the gas in the outflow. The density of the galaxy can be inferred from its rotational 
velocity $vel_{N4569}(r)$ $\simeq$ 250 km s$^{-1}$ at a radius $r$ = 8 kpc (Rubin et al. 1989) using the virial theorem ($\rho_{N4569}$ $\simeq$ 4.9 $\times$ 10$^{-24}$ g cm$^{-3}$), while 
that of the outflow is the one previously derived from the H$\alpha$ luminosity ($\rho_{out}$ $\simeq$ 1.4 $\times$ 10$^{-25}$ g cm$^{-3}$).
The potential energy of the outflow is thus $\Phi_{out}$ $\simeq$ 9 $\times$ 10$^{55}$ ergs.
The kinetic energy of the outflow is:

\begin{equation}
{E_{kin,out} = \frac{vel_{out}^2}{2}M_{out}}
\end{equation}

\noindent
where $M_{out}$ is the mass of the outflow ($M_{out}$ = 3.4 $\times$ 10$^7$ M$_{\odot}$, see Table \ref{masse}) and $vel_{out}$ its velocity 
corrected for the orientation of the galaxy on the plane of the sky ($vel_{out}$ $\simeq$ 260 km s$^{-1}$, see Table \ref{tab:ratios}). The kinetic energy of the outflow is 
$E_{kin,out}$ $\simeq$ 2.4 $\times$ 10$^{55}$ ergs and the total energy is $E_{tot,out}$ $\simeq$ 1.1 $\times$ 10$^{56}$ ergs. 
This energy would require 1.1 $\times$ 10$^5$ supernovae of energy 10$^{51}$ ergs, or a larger number if we assume a more realistic 1-10\%  energy transfer efficiency.
To provide this number of supernovae, the nuclear star cluster should have a mass $\gtrsim$ 2.4 $\times$ 10$^7$ M$_{\odot}$ and a star formation rate of $\gtrsim$ 24 M$_{\odot}$ yr$^{-1}$.
The required mass and star formation activity of the nuclear starburst are a few order of magnitudes higher than those observed 
in the galaxy (Keel 1996; Barth \& Shield 2000, Boselli et al. 2015). We caution, however, that ram pressure can 
produce low-density superbubble holes in the inner disc and thus supply extra energy to the outflow through Kelvin-Helmholtz instabilities (Roediger \& Hensler 2005)
and viscous stripping (Roediger \& Bruggen 2008). Despite this possible increase of the efficiency in the energy transfer to the outflow due to ram pressure,
this simple calculation suggests a probable suply of energy by an AGN.

The striking velocity difference between the adjacent regions G and H ($\sim$ 60 km s$^{-1}$) can be explained by the fact that while region G is associated to the nuclear outflow,
in region H the ionised gas might have been stripped from the disc of the galaxy and is only located in projection close to the outer extension of the outflow. Indeed, a
deep inspection of Fig. \ref{Netcore} shows that region H is a few arcsec below the projection of the nuclear outflow. 
 
\subsection{The diffuse gas}

The rough estimate of the {
total mass of the ionised gas derived from H$\alpha$ in 
the diffuse tail that we have derived in sect. 4 is $M_{tail}(H\alpha)$ $\simeq$ 3.2 $\times$ 10$^{9}$ M$_{\odot}$ (Table \ref{masse}). This mass can be 
compared to the total dynamical mass of the galaxy (1.2 $\times$ 10$^{11}$ M$_{\odot}$, Haan et al. 2008) 
and to the mass of gas in the other gaseous phases. This galaxy has a total mass of molecular hydrogen of $M(H_2)$ = 4.9 $\times$ 10$^9$ or
2.2 $\times$ 10$^9$ M$_{\odot}$ depending whether the molecular gas mass is derived from CO observations using a constant or variable CO-to-H$_2$ conversion factor
(Boselli et al. 2014b).  The total HI mass of NGC 4569 is 7.6 $\times$ 10$^8$ M$_{\odot}$ (Haynes et al. 2011) and is about a factor of ten smaller 
than the one of isolated galaxies of similar type and luminosity, as indeed indicated by its large HI-deficiency parameter
($HI-def$\footnote{The HI-deficiency parameter is defined as the difference in logarithmic scale between the expected and the observed HI
mass of a galaxy of given angular size and morphological  type (Haynes  \&  Giovanelli  1984).} = 1.05, Boselli et al. 2014b). 
This suggests that NGC 4569 has lost $\simeq$ 7.7 $\times$ 10$^9$ M$_{\odot}$ of HI. A more accurate estimate of the total mass of gas lost
by the galaxy during its interaction with the cluster environment can be derived from multizone chemo-spectrophotometric model of galaxy
evolution (Boselli et al. 2006). The truncated gaseous and stellar profiles of NGC 4569 can be reproduced if the galaxy lost 
$\simeq$ 1.9 $\times$ 10$^{10}$ M$_{\odot}$ of gas during a ram pressure stripping event that started $\sim$ 100 Myr ago. 
This mass is slightly larger than the very rough estimate of the ionised gas mass derived from H$\alpha$  
and suggests that a large fraction of the atomic gas, once stripped, is ionised within the tail. The deep VLA 21 cm observations of the VIVA survey detect
HI gas only within the stellar disc of NGC 4569 (Chung et al. 2009). A comparison of the HI and H$\alpha$+[NII] frames of the galaxy does
not show the presence of any HI extraplanar feature associated to the tails of ionised gas down to a column density limit of 
3-5 $\times$ 10$^{19}$ cm$^{-2}$ for a 10 km s$^{-1}$ spectral resolution (Chung et al. 2009). If all the stripped HI gas ($\simeq$ 1.9 $\times$ 10$^{10}$ M$_{\odot}$)
was still in its neutral phase and was distributed within the same tail defined by the H$\alpha$+[NII] emission, its column density  
should be $\Sigma(HI)$ $\simeq$ 0.4 M$_{\odot}$ pc$^{-2}$ ($\sim$ 5 $\times$ 10$^{19}$ at cm$^{-2}$) close to the detection limit
of the VLA images. This, however, is probably a lower limit since we expect that the gas, as indicated by the observations of the 
ionised phase or by the simulation of the HI phase (Tonnesen \& Bryan 2010) should have a clumpy distribution with peaks in column 
density well above this limit. The MUSE observations of ESO 137-001 show that the kinematical properties of the stripped gas do 
not significantly change with respect to that of the parent galaxy (Fumagalli et al. 2014). We thus expect that the stripped gas 
is distributed on the same velocity range than NGC 4569 and it is not spread on a much wider range, reducing the expected signal-to-noise per channel.
We can thus conclude that, if the stripped gas was still in its neutral phase, we should detect it.

An independent estimate of the upper limit to the column density of the neutral gas in the tail can be inferred using the deep \textit{Herschel} observations shown in Fig. \ref{XFIR}.
Considering the typical sky noise level at 250 $\mu$m around NGC 4569 given in Ciesla et al. (2012) ($\simeq$ 0.25 mJy pixel$^{-2}$ corresponding to $\simeq$ 0.07 mJy arcsec$^{-2}$),
we can derive the detection limit in dust column density using a modified black body emission with $\beta$ = 2 and a grain emissivity parameter $k_{250}$ = 2.0 cm$^2$ g$^{-1}$ 
(e.g. Boselli 2011). This limit can be transformed into a limit in gas column density using a typical gas-to-dust ratio. The gas-to-dust ratio of nearby massive galaxies ranges  
from 160 (Sodroski et al. 1994) to 70 (Sandstrom et al. 2013). With these values, the detection limit in the gas column density derived from 250 $\mu$m data should be
$\sim$ 2-5 $\times$ 10$^{19}$ cm$^{-2}$, comparable to the limits of the VIVA survey.

There are other indications suggesting that the gas is ionised within the tail. We can calculate the typical timescale necessary for the ionised gas 
in the tail to recombine using the relation:

\begin{equation}
{\tau_{rec} = \frac{1}{n_e \alpha_A}}
\end{equation}

\noindent
where $\alpha_A$ is the total recombination coefficient ($\alpha_A$ = 4.2 $\times$ 10$^{-13}$ cm$^3$ s$^{-1}$; Osterbrock \& Ferland 2006).
For a typical density of $n_e$ $\simeq$ 2-4 $\times$ 10$^{-2}$ cm$^{-3}$, the recombination time is $\simeq$ 2 Myr, 
a short time if compared to the time necessary to produce a tails $\simeq$ 145 kpc long (see sect. 4). It is thus conceivable that the gas is kept ionised within the tail.
The comparison of the H$\alpha$+[NII] frame and the optical and UV images in Fig. \ref{RGB}, \ref{NGVS}, and \ref{UV} shows the lack of any
compact star forming region within the tail ruling out in situ stellar photoionisation. The detailed comparison of the 
ionisation models with the spectroscopy properties of the gas in the tail of ESO 137-001 suggests that the gas is not only photoionised by the UV radiation emitted by
young stars but also by other mechanisms. These might be ionisation by the hot gas of the intracluster medium, thermal conduction and turbulent mixing, although a direct evidence of their presence 
is still lacking (Tonnesen et al. 2011; Fossati et al. 2015). The lack of any star forming region in the tail, 
on the contrary present in ESO 137-001 (Jachym et al. 2014), suggests that the contribution of these other mechanisms must be even dominant in NGC 4569.

The lack of star forming regions in the tail of NGC 4569 can result from two main effects as indicated by the simulations of Tonnesen \& Bryan (2012).
The efficiency with which the gas is transformed into stars in the tails of ram pressure stripped galaxies depends on the way the low-density
gas cools and condenses in the turbulent wakes. This process is more efficient whenever the density of the intracluster medium is high as 
in massive clusters such as Coma and A1367 where most of the tails of stripped material harbor star forming regions (Yoshida et al. 2008, 
Yagi et al. 2010, Fossati et al. 2012). The typical density of the intracluster medium in Virgo is $\sim$ a factor of 10 lower 
than in Coma and A1367 (Briel et al. 1992; B\"ohringer et al. 1994). Considering a $\beta$-model (Cavaliere \& Fusco-Femiano 1976) to trace the deprojected
distribution of the X-ray emitting gas within Virgo:

\begin{equation}
{\rho = \rho_0 \Big[1+(\frac{r}{r_c})^2\Big]^{-\frac{3}{2}\beta}}
\end{equation}

\noindent
and assuming a central density of the intergalactic medium of $\rho_0$ = 2 $\times$ 10$^{-3}$ cm$^{-3}$ (B\"ohringer et al. 1994), 
a core radius $r_c$ = 2.7 arcmin and $\beta$ = 0.47 (Schindler et al. 1999), we estimate that the density of the intracluster medium 
near NGC 4569 is $\rho$ $\simeq$ 10$^{-5}$ cm$^{-3}$ (we assume a distance $r$ from the cluster core of 1.7 degrees, corresponding to 0.32 $R_{vir}$). 
This value can be compared to the density of the gas used in the simulation of Tonnesen \& Bryan (2012), 
$\rho$ $=$ 5 $\times$ 10$^{-5}$ cm$^{-3}$ for a galaxy of stellar mass 10$^{11}$ M$_{\odot}$ (vs. $M_{star}$ = 3 $\times$ 10$^{10}$ M$_{\odot}$ for NGC 4569).
The simulations of Tonnesen \& Bryan (2012) also indicate that for the star formation to take place 
in the tail requires a sufficient amount of time ($\sim$ 200 Myr) to make the gas cool and collapse. This timescale is comparable to the 
derived age of the interaction (100-300 Myr; Vollmer et al. 2004, Boselli et al. 2006; Crowl \& Kenney 2008). It is thus possible that the 
stripped gas still did not have the time to collapse and form new stars. We should recall, however, that these condensed regions of star formation 
appear in the simulations whenever the gas is allowed to cool and the contribution of the different heating processes is underestimated. The presence of 
several condensed regions in another Virgo cluster galaxy, IC 3418 (Hester et al. 2010; Fumagalli et al. 2011; Kenney et al. 2014), 
where the conditions of the intracluster medium are expected to be similar 
to those encountered by NGC 4569, clearly indicate that the process of formation of these extraplanar HII regions is still far from being understood.

\subsection{The evolution of NGC 4569 in the cluster}

The new set of extremely deep H$\alpha$+[NII] imaging data collected in this work, combined with those available at other frequencies, 
and the comparison with model predictions allow us to reconstruct the evolution of the galaxy within the Virgo cluster environment. 
All observational evidence collected so far suggests that NGC 4569 underwent a recent ram pressure stripping event.
Given the presence of a nearby companion, IC 3583, at a projected distance of $\sim$ 6 arcmin, 
however, we cannot exclude gravitational perturbations. A gravitational interaction between the two objects, however, seems ruled out by the fact
that their most accurate distance estimate done using the tip of the red giant branch in HST observations locates IC 3583 in the foreground of the cluster ($D$ = 9.52 Mpc)
and NGC 4569 at $D$ $\gtrsim$ 17 Mpc (Karachentsev et al. 2014).
To quantify the importance of any possible tidal interaction with that galaxy we can estimate the duration of a possible tidal encounter between the two galaxies
using the relation (Binney \& Tremaine 1987):

\begin{equation}
{t_{enc} \simeq max[r_{NGC4569}, r_{IC3583}, b]/\Delta V}
\end{equation}

\noindent
where $r_{NGC4569}$ (22.8 kpc) and $r_{IC3583}$ (6.9 kpc) are the radii of the two galaxies, $b$ their separation, and $\Delta V$ their relative velocity. 
If we assume that both galaxies are at the same distance, thus that $b$ $\simeq$ 32 kpc we can calculate $t_{enc}$ using eq. 7. IC 3583 has a radial velocity of 1120 km s$^{-1}$,
significantly different from that of NGC 4569 ($vel$ = -221 km s$^{-1}$), thus $t_{enc}$ $\simeq$ 23 Myr. This timescale is very short compared to the time required 
to NGC 4569 to make a complete revolution ($\simeq$ 370 Myr). Although an interaction on their extended halos is still possible, it is quite
unlikely that on such short 
timescale the tidal interaction is able to remove $\simeq$ 1.9 $\times$ 10$^{10}$ M$_{\odot}$ of atomic gas.

We can also estimate the typical truncation radius for two interacting galaxies using the relation (Read et al. 2006):

\begin{equation}
{r_t \simeq b \Big[\frac{m}{M(3+e)}\Big]^{1/3}}
\end{equation} 

\noindent
where $m$ and $M$ are the masses of the two interacting galaxies, $b$ their separation and $e$ the ellipticity of their orbit. Again assuming $b$
= 32 kpc as a lower limit, an ellipticity of $e$ = 1 and a stellar mass $M_{star}$ = 3 $\times$ 10$^{10}$ M$_{\odot}$ for NGC 4569
and $M_{star}$ = 6.3 $\times$ 10$^{8}$ M$_{\odot}$ for IC 3583 we obtain a truncation radius $r_t$ $\gtrsim$ 74 kpc for NGC 4569 and 5.5 kpc for IC 3583.
The truncation radius of NGC 4569 is significantly larger than the optical radius (22.8 kpc), it is thus very unlikely that gravitational interactions 
have been the dominant perturbing mechanism affecting the recent evolution of that galaxy. 
This result is consistent with the fact that we do not observe any significant stellar tidal feature associated 
to the observed tails of ionised gas (Fig. \ref{RGB}). We cannot, however, exclude that a fly-by encounter of the two galaxies has occurred.
This kind of encounter generally induces nuclear gas infall (Moore et al. 1998). If this happened, it could explain the prodigious nuclear
starburst activity occurred 3-6 Myr ago (Barth \& Shields 2000; Grebel \& Bruhweiler 2002), or an older ($>$ 15 Myr) starburst (Keel 1996),
with possible feeding of a mild nuclear activity. The spectacular optical images obtained at the CFHT as part of the NGVS survey (Fig. \ref{NGVS}) indicate
small (a few kpc) low surface brightness features perpendicular to the disc in the south east side of the galaxy or in the north suggesting a possible bridge 
with IC 3583. A minor gravitational interaction might also have contributed to flatten the potential well of the two galaxies, weakening their gravitational binding forces 
which keep the diffuse gas of the ISM anchored to the stellar disc, thus making ram pressure stripping more efficient 
(Gavazzi et al. 2001).

A further evidence in favour of a ram pressure stripping event is the presence of a polarised radio ridge southwest to the galaxy centre 
probably produced by a local compression of the gas able to organise the magnetic field (Wezgowiec et al. 2012). This polarised radio continuum feature is 
located south of the main spur of ionised gas coming out from the nucleus of the galaxy. The position of the 
tails of ionised gas suggests a slightly different orbit than the one proposed by Vollmer (2009), where the galaxy is expected to have crossed the 
cluster core $\sim$ 300 Myr ago and is now coming towards us on a south-west to north-east orbit. The observed tails of H$\alpha$+[NII] gas 
rather indicate a west-to-east orbit, questioning thus the interpretation of the X-ray distribution of the hot gas in a Mach cone
as proposed by Wezgowiec et al. (2011).

The first interesting result of this work is that ram pressure stripping can be the dominant mechanism for removing the ISM 
in massive galaxies ($M_{star}$ $\simeq$ 3 $\times$ 10$^{10}$ M$_{\odot}$) falling into intermediate class
clusters ($M_{virial}$ = 1.4-4.2 $\times$ 10$^{14}$ M$_{\odot}$, McLaughlin 1999, Urban et al. 2011, Nulsen \& Bohringer 1995, 
Schindler et al. 1999; $\Delta_{vel}$ = 800 km s$^{-1}$, Boselli et al. 2014a; $\rho_0$ = 2 $\times$ 10$^{-3}$ cm$^{-3}$
$T$ =2.3 keV, Bohringer et al. 1994), thus extending previous finding to a much broader range of environments 
and objects (see also Catinella et al. 2013). If the galaxy has fallen into the cluster from behind, as indicated by its negative recessional velocity, in an orbit from west to
east, as suggested by the tails, it has encountered the maximal density of the intracluster gas north to M87, at a radial distance of $\gtrsim$ 230 kpc ($\simeq$ 0.1
$R_{vir}$), where the density of the intergalactic medium is $\rho_{230 kpc}$ $\simeq$ 4 $\times$ 10$^{-5}$ cm$^{-3}$ as derived using equation 4.
It is worth mentioning that NGC 4569 is not an isolated case of massive galaxy with tails of gas witnessing an ongoing ram pressure stripping event
in the Virgo cluster:
NGC 4388 is another obvious candidate (Yoshida et al. 2002; Oosterloo \& van Gorkom 2005; Kenney et al. 2008), as well as the seven massive spirals 
with HI tails observed by Chung et al. (2007).

This analysis also shows that the diffuse component of the ionised gas in the extended tail is a factor of $\sim$ 90 larger 
than that expelled by the nuclear outflow. Since the ionised gas is the dominant phase in the tail (the mass of hot gas in the galaxy halo derived from X-ray data is
$\lesssim$ 2 $\times$ 10$^8$ M$_{\odot}$; Wezgowiec et al. 2011), this indicates that in massive galaxies the contribution of the nuclear feedback to the 
ejection of the gas mass is minimal. Furthermore it can hardly reproduce the truncated disc in the gas and dust components observed in NGC 4569 and in most of the 
gas deficient cluster galaxies with a quenched activity of star formation (Cortese et al. 2012a; Boselli et al. 2014c). 
As suggested by hydrodynamic cosmological simulations, however, it can contribute to make ram pressure stripping efficient by injecting kinetic energy into the ISM
weakening the gravitational forces which keep the gas bounded to the potential well of the galaxy (Bahe \& McCarthy 2015).
The question is whether NGC 4569 is representative of typical massive galaxies in terms of nuclear activity. 
It is classified as a LINER/HII region transition type nucleus by Gabel \& Bruhweiler (2002). It is also classified as a strong AGN using the BPT
diagram (Baldwin et al. 1981), as  17 \% of the \textit{Herschel} Reference Survey late-type galaxies with a stellar mass $M_{star}$ $>$ 10$^{10}$ M$_{\odot}$ (Gavazzi et al., in preparation),
and can thus be considered as a typical active massive galaxy. The feedback process that follows the removal of the hot X-ray halo of galaxies 
falling in high-density environments in a starvation scenario (Larson et al. 1980) does not seem to be as efficient as cosmological simulations 
or semi-analytic models indicate (Weinmann et al. 2006; McCarthy et al. 2008, 2011; Font et al. 2008; Kang \& van den Bosch 2008; McGee et al. 2009; 
Kimm et al. 2009; Guo et al. 2011; De Lucia et al. 2012; Bahe \& McCarthy 2015). A more realistic description of the (nuclear) feedback 
process should be considered. 
Our results provide further evidence that ram pressure is a compelling mechanism to explain the stripping of the cold gas component of the ISM 
and thus the quenching of the star formation activity of late-type galaxies in high density environments.

\section{Conclusion}

We present new deep, narrow-band H$\alpha$+[NII] imaging data of NGC 4569 obtained with MegaCam at the CFHT. The H$\alpha$+[NII]
image show the presence of long low surface brightness ($\Sigma(H\alpha +[NII])$ $\simeq$ 10$^{-18}$ erg s$^{-1}$ cm$^{-2}$ arcsec$^{-2}$) 
tails of ionised gas extending perpendicularly from the disc of the galaxy in the west direction up to $\simeq$ 145 kpc. The presence of these
tails are a clear indication that the galaxy is undergoing a ram pressure stripping event. This observational evidence suggests that the ram pressure stripping mechanism is efficient 
not only in intermediate-to-low mass galaxies in the core of massive clusters, as previously thought, but also in massive galaxies located in an unrelaxed cluster 
of intermediate mass ($\sim$ 10$^{14}$ M$_{\odot}$) still in formation, with characteristics 
similar to those encountered in high-density regions at high redshift. 
The H$\alpha$+[NII] image also shows a plume of ionised gas extending 
$\simeq$ 8 kpc perpendicular to the nucleus powered by a nuclear outflow. The total mass of the ionised gas in the tail is an important fraction of that of the cold atomic hydrogen
that the galaxy has lost during its crossing of the cluster. The mass of the ionised gas expelled by the nuclear outflow, on the contrary,
is $\simeq$ 1 \% of the total mass of the ionised gas in the tail. It can hardly be at the origin of the truncated gaseous, dust and star forming disc of NGC 4569. 
If we consider NGC 4569 representative of massive galaxies in intermediate density regions, this 
analysis suggests that ram pressure stripping is the dominant process responsible for the gas removal and for the quenching of the star formation activity
observed in galaxies located in high-density regions. The contribution of the nuclear feedback, made efficient after the removal
of the hot gas halo (starvation), is only marginal and significantly less important than what it is generally assumed in cosmological simulations. \\

The lack of HII regions, the derived density and the physical extension of the tails let us speculate that the gas is mainly excited by 
mechanisms other than photoionisation. These can be shocks in the turbulent gas, MHD waves, and heat conduction. The presence of thin filamentary structures 
observed along the tails, reproduced by the most recent hydrodynamic simulations, suggests that magnetic fields might play an important role. \\

The analysis presented in this work underlines once more how the detailed study of representative objects in the nearby universe, where high-quality multifrequency data and 
tuned chemo-spectrophotometric and kinematic models are available, is a powerful tool for understanding the environmental mechanisms affecting galaxy evolution.
This work is a further evidence that deep narrow band H$\alpha$+[NII] imaging obtained with wide field detectors is probably the most sensitive technique to 
catch ongoing interactions such as the one observed in NGC 4569. Indeed, at the typical depth that modern instruments can provide, the fraction of galaxies in nearby clusters
with tails of stripped material is very small in HI or X-ray, while it strongly increases in H$\alpha$. As an example, the number of galaxies with HI tails in the VIVA survey
of the Virgo cluster is only 7 out of the 53 observed objects. For comparison, $\sim$ 50 \% of the late-type galaxies observed with a narrow band H$\alpha$+[NII] 
filter in Coma and A1367 by Yagi, Yoshida and collaborators have extended tails of ionised gas (e.g. Boselli \& Gavazzi 2014).  
The very nature of the physical process responsible for the stripping of the gas (ram pressure vs. tidal interactions), for its excitation 
in the tails or in the plume associated to the nuclear outflow and for the possible formation of HII regions far from the galactic disc, however, requires deep 
high velocity and angular resolution integral field spectroscopic observations that only instruments such as MUSE can provide. 

\begin{acknowledgements}

This research has been financed by the French ANR grant VIRAGE and the French national program PNCG.
We wish to thank the GALEX Time Allocation Committee for the generous allocation of time devoted to this project
and the anonimous referee for constructive comments.
M.Fossati acknowledges the support of the Deutsche Forschungsgemeinschaft via Project ID 3871/1-1. 
LC acknowledges financial support from the Australian Research Council (DP150101734).
M.Fumagalli acknowledges support by the Science and Technology 
Facilities Council [grant number  ST/L00075X/1].
E. Toloba is supported by the NSF grant AST-1412504.
This research has made use of the NASA/IPAC Extragalactic Database (NED) 
which is operated by the Jet Propulsion Laboratory, California Institute of 
Technology, under contract with the National Aeronautics and Space Administration
and of the GOLDMine database (http://goldmine.mib.infn.it/) (Gavazzi et al. 2003).

\end{acknowledgements}

\begin{table*}
\caption{H$\alpha$+[NII] fluxes and equivalent widths of Virgo cluster galaxies with data in the literature. }
\label{gal}
{
\[
\begin{tabular}{cccccc}
\hline
\noalign{\smallskip}
\hline
		& This work		& This work		& Literature		& Literature	\\
      Galaxy  &  log f(H$\alpha$+[NII])  & H$\alpha$+[NII]E.W. &    log f(H$\alpha$+[NII])  & H$\alpha$+[NII]E.W. &  Ref. \\
              & erg s$^{-1}$ cm$^{-2}$  &   \AA                & erg s$^{-1}$ cm$^{-2}$     &   \AA               &       \\      
\hline
NGC 4569      & -11.85$\pm$0.02		& 3.1$\pm$0.1		& -12.02$\pm$0.25	&	2$\pm$1		& 1\\
              &				& 			& -11.24		&	-		& 2\\
	      &				&			& -11.62$^b$		&	6$\pm$2$^b$	& 3\\
	      &				&			& -11.95$\pm$0.07	&	-		& 4\\
	      &				&			& -11.83		&	-		& 5\\
IC 3583       & -12.67$\pm$0.02		& 12.1$\pm$0.4		& -12.17$\pm$0.08	&   43.5$\pm$12.2	& 6\\
NGC 4584      & -13.07$\pm$0.04		& 4.8$\pm$0.3		& -12.98$\pm$0.19	&    7.5$\pm$4.8	& 6\\
AGC 225847    & -14.42$\pm$0.08		& 17.3$\pm$1.7		& -14.30$\pm$0.07	&   26.2$\pm$4.5	& 7\\
VCC 1614$^a$  & -14.70$\pm$0.01		& 8.4$\pm$0.1		& -14.75$\pm$0.04	&    4.3$\pm$0.5	& 8\\
\noalign{\smallskip}
\hline
\end{tabular}
\]
Note: a: in the 3 arcsec SDSS aperture; b: measured in a circular aperture of 7 arcmin diameter.\\
References: 1) Boselli \& Gavazzi (2002); 2) Young et al. (1996); 3) Kennicutt \& Kent (1983); 4) Sanchez-Gallego et al. (2012);
5) Koopmann et al. (2001); 6) Gavazzi et al. (2002); 7) Gavazzi et al. (2012); 8) Alam et al. (2015).
}
\end{table*}

\begin{table*}
\caption{Spectroscopic emission line measurements of the different regions shown in Figures \ref{Netcore} and \ref{2D}. 
}
\label{tab:ratios} 

\centering

\begin{tabular}{l r r c r r c c c}
\hline
\hline
   Region &  Start &  End  &  $\log$(H$\alpha$/H$\alpha_{nuc}$)   &  $\Delta v$ 	&  $\sigma$	     &    [NII]$\lambda$6584/H$\alpha$ &    [SII]$\lambda\lambda$6716,6731/H$\alpha$  &   [SII]$\lambda$6716/[SII]$\lambda$6731 \\
          &   "    &  "    &                                      &  $\rm (km~s^{-1})$  &  $\rm (km~s^{-1})$ &  			       &					      & 					\\
\hline  

      A   &  16    &	 7 &   -1.9  &   82.6$\pm$5.5	&    54.2$\pm$6.1  & 1.39$\pm$0.41 &   0.88$\pm$0.28 &   1.48$\pm$0.43 \\
      Nuc &   3    &	-3 &   +0.0  &    3.2$\pm$2.0	&   132.8$\pm$2.1  & 1.29$\pm$0.04 &   0.67$\pm$0.02 &	 1.10$\pm$0.06 \\
      B   &  -9    &   -14 &   -1.7  &   24.4$\pm$1.5	&    29.6$\pm$1.8  & 0.69$\pm$0.05 &   0.44$\pm$0.04 &	 1.40$\pm$0.25 \\
      C   & -21    &   -33 &   -2.7  &   10.6$\pm$8.4	&    41.6$\pm$8.8  & 2.15$\pm$0.80 &         -	     &	       -       \\
      D   & -49    &   -54 &   -1.8  &   60.0$\pm$0.8	&     9.6$\pm$1.7  & 0.37$\pm$0.03 &   0.26$\pm$0.02 &	 1.74$\pm$0.32 \\
      E   & -67    &   -84 &   -2.2  &  128.4$\pm$7.8	&    89.2$\pm$8.2  & 0.85$\pm$0.12 &   0.48$\pm$0.08 &	 1.20$\pm$0.30 \\
      F   & -84    &   -89 &   -2.4  &   84.1$\pm$11.4  &    73.8$\pm$12.8 & 0.79$\pm$0.21 &         -	     &	       -       \\
      G   & -89    &   -96 &   -2.1  &   48.8$\pm$4.8	&    79.1$\pm$4.9  & 1.06$\pm$0.11 &   0.77$\pm$0.09 &	 1.51$\pm$0.32 \\
      H   & -96    &  -104 &   -2.4  &  -13.9$\pm$8.6	&    90.6$\pm$9.5  & 1.30$\pm$0.22 &         -       &	       -       \\
\hline
\end{tabular}
Notes: Region Nuc corresponds to the nucleus of NGC 4569. Start and End denote the position in arcseconds with respect to the 
photometric center where the 1D extraction is performed. $\log$(H$\alpha$/H$\alpha_{nuc}$) is the brightness of the H$\alpha$
line in a given region normalised to that of the nucleus. $\Delta_V$ is given with respect to the systemic
velocity of NGC 4569 ($-221 \rm{km s^{-1}}$).
\end{table*}

\begin{table*}
\caption{H$\alpha$+[NII] flux and surface brightness of the low surface brightness features. }
\label{LSB}
{
\[
\begin{tabular}{cccc}
\hline
\noalign{\smallskip}
\hline
Region  &  log f(H$\alpha$+[NII]) & $\Sigma(H\alpha +[NII]) $	                    & Area         \\
        & erg s$^{-1}$ cm$^{-2}$  & 10$^{-19}$ erg s$^{-1}$ cm$^{-2}$ arcsec$^{-2}$ & arcsec$^2$   \\      
\hline
Tail	& -12.52$\pm$0.36	&   5.2$\pm$4.4		& 573619\\
Outflow & -13.07$\pm$0.02	&  88.2$\pm$4.4 	& 9625	\\
1	& -14.19$\pm$0.13	&  14.4$\pm$4.4		& 4537  \\
2	& -14.19$\pm$0.18	&  10.5$\pm$4.4		& 6185  \\
3	& -14.14$\pm$0.20	&   9.3$\pm$4.4		& 7743  \\
4	& -14.22$\pm$0.18	&  10.7$\pm$4.4		& 5628	\\
5	& -14.35$\pm$0.16	&  11.8$\pm$4.4		& 3773	\\
6	& -14.59$\pm$0.18	&  10.7$\pm$4.4		& 2404	\\
7	& -14.40$\pm$0.15	&  12.4$\pm$4.4		& 3251	\\
8	& -13.41$\pm$0.12	&  16.2$\pm$4.4		& 24350	\\
9	& -13.61$\pm$0.16	&  11.9$\pm$4.4		& 20417	\\
10	& -13.39$\pm$0.09	&  21.2$\pm$4.4		& 19146	\\
11	& -13.90$\pm$0.15	&  12.3$\pm$4.4		& 10220	\\
12	& -14.12$\pm$0.17	&  11.4$\pm$4.4		& 6667	\\
13	& -14.26$\pm$0.17	&  11.0$\pm$4.4		& 5038	\\
\noalign{\smallskip}
\hline
\end{tabular}
\]
}
\end{table*}

\begin{table*}
\caption{Parameters of NGC 4569 and IC 3583 }
\label{masse}
{
\[
\begin{tabular}{ccccc}
\hline
\noalign{\smallskip}
\hline
Variable				&  NGC 4569 		& Ref. &	IC 3583 & Ref.   \\
     
\hline
Morph. type				& SAB(rs)ab;LINER;Sy	& 1 & IBm		& 1	\\
$D$  (Mpc)				& $\gtrsim$ 17		& 2 & 9.52$\pm$0.95	& 2	\\
$r_{25}$				& 22.8	kpc		& 3 & 6.9       	& 4	\\
$vel$					& -221	km s$^{-1}$	& 5 & 1120		& 5	\\ 
$r$/$R_{vir}$				& 0.32			& TW& 0.32              & TW    \\
$M_{star}$ (M$_{\odot}$)		& 3.0$\times$10$^{10}$ 	& 6 & 6.3$\times$10$^{8}$& 6	\\
$M_{dyn}$ (M$_{\odot}$)		    	& 1.2$\times$10$^{11}$ 	& 7 & -                 & 	\\

$M(HI)$ (M$_{\odot}$)			& 7.6$\times$10$^{8}$ 	& 5 & 4.7$\times$10$^{8}$& 5	\\
$M(H_2)$$^a$ (M$_{\odot}$)		& 4.9$\times$10$^{9}$ 	& 8 &$<$2.8$\times$10$^{7}$& 9	\\
$M(X-rays)$ (M$_{\odot}$)		& $\lesssim$ 2$\times$10$^{8}$	& 10 & - & - \\		
$M_{tail}(H\alpha)$ (M$_{\odot}$)	& 3.2$\times$10$^{9}$ 	& TW& -			& - \\
$M_{out}(H\alpha)$ (M$_{\odot}$)	& 3.4$\times$10$^{7}$ 	& TW& -			& - \\
\noalign{\smallskip}
\hline
\end{tabular}
\]
Notes: $a$ = derived using a constant $X_{CO}$ = 2.3 $\times$ 10$^{20}$ cm$^{-2}$/(K km s$^{-1}$) conversion factor (Boselli et al. 2002).\\
References: 1) NED; 2) Karachentsev et al. (2014); 3) $g$-band isophotal radius, from Cortese et al. (2012b); 4) GOLDMine (Gavazzi et al. 2003); 5) Haynes et al. (2011);
6) derived from $i$-band luminosities using the $g-i$ colour-dependent stellar mass-to-light ratio relation from Zibetti et al. (2009) 
and assuming a Chabrier (2003) initial mass function; 7) Haan et al. (2008); 8) Boselli et al. (2014b); 9) Boselli et al. (2002); 10) Wezgowiec et al. (2011).\\
}
\end{table*}

\end{document}